\documentclass[twocolumn,aps,prd,a4paper,final,superscriptaddress,longbibliography,showpacs,showkeys,nofootinbib]{revtex4-2}

\usepackage{amsmath,amssymb}
\usepackage{ascmac}
\usepackage{mathrsfs}
\usepackage[T1]{fontenc}
\usepackage{physics}
\usepackage{bm}
\usepackage{braket}
\usepackage{slashed}
\usepackage{float}
\usepackage{natbib}
\usepackage{mathtools}
\usepackage{xcolor}
\usepackage{comment}
\usepackage{subcaption}
\usepackage[normalem]{ulem} 

\usepackage[pdftex]{graphicx}
\usepackage{hyperref}

\begin{document}

\newcommand{\ctext}[1]{\raise0.2ex\hbox{\textcircled{\scriptsize{#1}}}}
\renewcommand{\figurename}{Fig. }
\renewcommand{\thefigure}{\arabic{figure}}
\renewcommand{\tablename}{Table.}
\renewcommand{\thetable}{\arabic{table}}

\newcommand{\txtr}[1]{\textcolor{red}{#1}}
\newcommand{\txtb}[1]{\textcolor{blue}{#1}}
\newcommand{\txtc}[1]{\textcolor{cyan}{#1}}
\newcommand{\txtm}[1]{\textcolor{magenta}{#1}}

\newcommand{\ya}[1]{\textcolor{red}{#1}}
\newcommand{\sy}[1]{\textcolor{blue}{#1}}
\newcommand{\ns}[1]{\textcolor{cyan}{#1}}

\newcommand{\Erase}[1]{{\color{red}\sout{\color{black}{#1}}}%
}

\newcommand{\CP}{\mathbb{C}\mathrm{P}}

\newcommand{\yamamotoo}[1]{\textcolor{purple}{#1}}

\twocolumngrid


\title{Topological Charge Asymmetry in a $\CP^N$ Skyrmion-Fermion Coupled System}



\author{Yuki Amari}
\affiliation{Research and Education Center for Natural Sciences, Keio University, Hiyoshi 4-1-1, Yokohama, Kanagawa 223-8521, Japan}
\affiliation{Department of Physics, Keio University, 4-1-1 Hiyoshi, Kanagawa 223-8521, Japan}

\author{Nobuyuki Sawado}
\affiliation{Department of Physics and Astronomy, Faculty of Science and Technology, Tokyo University of Science, Noda, Chiba 278-8510, Japan}

\author{Shintaro Yamamoto}
\affiliation{Department of Physics and Astronomy, Faculty of Science and Technology, Tokyo University of Science, Noda, Chiba 278-8510, Japan}


\vspace{.5 in}
\small

\keywords{Topological Charge, $\CP^N$ Skyrmion, Dirac Fermion, Index Theorem}
\date{\today}

\begin{abstract}
	Topology plays a central role in classifying solitonic configurations in field theories, providing robustness and a nonperturbative label, the so-called topological charge $Q$. In soliton-fermion coupled systems, the relation between the topological charge and the number of zero modes is well established through the index theorem. However, the physical consequences of the sign of the topological charge have remained largely unexplored. In this work, we study fermions in $2+1$ dimensions coupled to Skyrmions with target space $\CP^N$, particularly focusing on the backreactions of the fermions and on the sign of the topological charge. We obtain the solutions in a self-consistent manner, which exhibit an asymmetry with respect to the topological charge $\pm Q$ especially in the strong coupling regimes. This asymmetry is caused from the fermionic eigenvalue problem inherent in the self-consistent formulation. Although the Lagrangian is symmetric under $Q\to-Q$, the coupled equations for the Skyrmions and anti-Skyrmions become inequivalent once fermionic backreaction is taken into account. We demonstrate the mechanism in $\CP^1$ and $\CP^2$ Skyrmions, but the analysis is directly extendable for the general $\CP^N$.
\end{abstract}

\maketitle 


\section{Introduction}
\label{Introduction}
Topological solitons are localized, nonperturbative field configurations that appear in a wide class of nonlinear systems~\cite{Rajaraman:1982is,Manton:2004tk,Manton:2022fcb,RevModPhys.51.591,Shnir:2018yzp}. They arise in diverse physical contexts, ranging from vortices and spin textures in condensed matter to nonperturbative excitations in relativistic field theories, such as monopoles and instantons in gauge theories and solitonic models of baryons in effective descriptions of Quantum chromodynamics (QCD). In general, topological solitons can be formulated as mappings from physical space to a target manifold, and the corresponding homotopy classes define an integer-valued quantity, the topological charge $Q$. This quantity is invariant under continuous deformations of the field and thereby ensures the stability of localized configurations. 

\begin{figure}[t]
	\centering
		\includegraphics[width=1\linewidth]{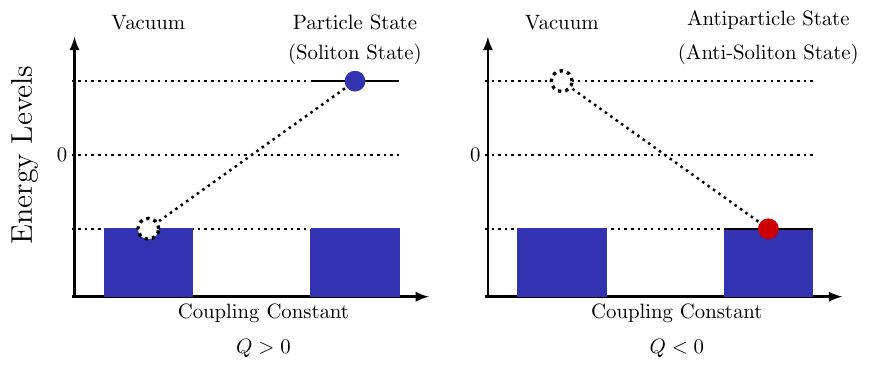}
		\caption{Schematic illustration of a spectral flow with solitons $(Q>0)$ and anti-solitons $(Q<0)$.}
		\label{QQ}
\end{figure}

The topological charge $Q$ not only guarantees the stability of solitons but also plays a crucial role in the localization of fermions in soliton-fermion coupled systems~\cite{Jackiw:1975fn,Gousheh:2012pwg,Jackiw:1981ee,Amari:2019tgs,Callan:1982ac,Rubakov:1988aq,Kahana:1984be,Kahana:1984dx,Jackiw:1977pu,Kiskis:1978tb,Kunz:1994ah}. The Dirac operator admits normalizable zero energy eigenstates, so-called zero modes, and the net number of such modes corresponds to the topological charge of the background soliton~\cite{Weinberg:1981eu,Abanov:2001iz,Abanov:2004ci}. This correspondence is supported by the Atiyah-Patodi-Singer index theorem, providing a connection between a topological quantity and analytical spectra of matters~\cite{Atiyah:1963zz,atiyah_patodi_singer_1975}. This feature can be investigated in more detail through the spectral flow analysis as illustrated in Fig.~\ref{QQ}, where the rearrangement of the fermion spectra under continuous deformation of the background field illustrates how fermionic energy levels cross the zero energy line~\cite{Kahana:1984be,Kahana:1984dx,Niemi:1984vz,Burnier:2006za,Kodama:2008xm,Delsate:2011aa,Amari:2019tgs,Amari:2023gjq,Amari:2024rpm}. 

In recent years, increasing attention has been paid to going beyond the background approximation by explicitly incorporating the backreaction of fermions on the soliton~\cite{Gani:2010pv,Amado:2014waa,Blas:2022oxs,Perapechka:2018yux,PhysRevD.100.105003,PhysRevD.101.021701,Perapechka:2019upv,Gani:2022ity,PhysRevD.108.065005,Dzhunushaliev:2024kti,PhysRevD.111.084072,Amari:2025rgt}. In this approach, the Dirac equation and the soliton field equation are solved simultaneously or self-consistently, allowing one to investigate how fermion localization affects the soliton configuration itself. It has been demonstrated that the presence of fermionic zero modes or bound states 
induce a remarkable deformation of the soliton profile. This self-consistent treatment makes it possible to uncover properties of solitons that are invisible in the background approximation. 

Despite these developments, an important issue about the sign of topological charge $Q$ remains unresolved. The previous studies have focused on configurations with either positive or negative topological charge. From a physical viewpoint, this raises a fundamental and still insufficiently explored question: can the sign of the topological charge have observable consequences beyond its conventional role as a classification label for soliton and antisoliton sectors? If the sign of $Q$ are to influence how a soliton responds to external fields or self-consistent interactions, it would indicate that the sign itself carries physical information. Although the energy of a soliton and an anti-soliton degenerates in many cases, not all systems possess this degeneracy; for example, it is lifted in chiral magnets~\cite{Nagaosa:2013ftn,Koshibae:2016aa,Hoffmann:2017,GOBEL20211} and in QCD with a strong magnetic field~\cite{Amari:2024adu}. Once the backreaction from fermions is included, similar inequivalence between solitons $\qty(Q>0)$ and anti-solitons $\qty(Q<0)$ is expected to appear. Since the fermion deforms the soliton configuration, the whole system is completely different in the sign of $Q$. In this paper, we refer the inequivalence as topological charge asymmetry\footnote{Although such an inequivalence has been referred to as a soliton-antisoliton asymmetry~\cite{Amari:2024adu}, we use the term ``topological charge asymmetry'' to highlight that the effect originates from the fermionic spectral properties associated with the sign of the topological charge as shown at the end of this paper. }.

In this work, we address this issue by studying the fermion backreaction on a Skyrmion with the target space $\CP^N$ in $2+1$ dimensions, which we simply call $\CP^N$ Skyrmion~\cite{PhysRevLett.100.047203,Ferreira:2010jb,Amari:2016ynl,Akagi:2021lva,Zhang:2022lyz,Amari:2022boe,Amari:2024pnw,PhysRevD.92.045007} as an example. The Skyrmions were originally proposed by Skyrme in the early 1960s as a nonlinear theory of the pion field, and they provided an effective model of baryons by identifying the baryon number with the topological charge~\cite{Skyrme:1961vq,Skyrme:1961vr}. The configurations considered in the present work are lower-dimensional analogues of these Skyrmions, commonly referred to as baby Skyrmions. While retaining the essential topological features, they allow for a more transparent analysis of fermion–soliton interactions. When $\CP^N$ Skyrmions possess rotational symmetry, the windings and the boundary conditions can be chosen independently, and their combination determines the sign of the topological charge. Consequently, the sign of the topological charge can be reversed by changing the winding orientation alone, without modifying the boundary condition. This feature is crucial for our purpose, as it enables a controlled comparison between configurations with opposite topological charge under the same boundary conditions, which is essential for identifying intrinsic topological charge asymmetries induced by fermionic backreaction. In practice, this separation is conveniently realized within an rotationally symmetric configuration. Moreover, fermions coupled to two-dimensional Skyrmions provide effective models for Dirac fermions interacting with topological textures or defects in condensed matter systems.

In the present work, we focus on $N=1,2$, treating the $\CP^1$ Skyrmions as trivially embedding configurations within the $\CP^2$ framework. We find that the topological charge asymmetry appears in both cases, indicating that the effect is not tied to special features of a particular target space, but rather originates from the generic interplay between fermions, topology, and backreaction. 

This paper is organized as follows. In Sec.~\ref{TheModel}, the $\CP^N$ Skyrmion-fermion coupled model is introduced. Moreover, we write down equations of motion with rotationally symmetric ansatz. Then, we also discuss how the $\CP^1$ Skyrmion can be realized as a specific configuration in the $\CP^2$ model. In Sec.~\ref{NumericalMethod}, we explain the boundary conditions, and the numerical computation method. Typical numerical results, which reveal the topological charge asymmetry, are presented in Sec.~\ref{main}. In Sec.~\ref{Mechanism}, the mechanism of the asymmetry is discussed. We conclude with a summary and outlook in Sec.~\ref{SummaryandOutlook}.



\section{The Model and Ansatz}
\label{TheModel}
We consider the $\CP^N$ Skyrmion-Dirac fermion coupled system in $2+1$ dimensions in the form
\begin{align}
	\mathcal{L}
	&=\overline{\Psi}\qty(i\hat{\gamma}^\mu\partial_\mu-m\hat{\Sigma})\Psi\nonumber\\
	&+g^{-2}\qty[\frac{M^2}{2}\mathrm{Tr}\qty(\partial_\mu \Sigma\partial^\mu \Sigma)
	{+e^2\mathrm{Tr}\qty(\qty[\partial_\mu \Sigma,\partial_\nu \Sigma]^2)}
	-V],
	\label{FullLagrangian}
\end{align}
with $M^2, e^2, g^{-2}>0$. Here, $\Psi$ represents a spinor multiplet and $\Sigma$ is a Hermitian matrix-valued field\footnote{The field $\Sigma$ has a linear relation with one of the $\mathrm{SU}\qty(N+1)$ color field $\mathfrak{n}$, such that $\mathfrak{n}=\frac{N-1}{N+1}I_{N+1}-\Sigma$. } 
defined as 
\begin{equation}
    \Sigma=I_{N+1}-2Z\otimes Z^\dagger \,
\end{equation} 
with the $\qty(N+1)\times\qty(N+1)$ unit matrix $I_{N+1}$ and the $\qty(N+1)$-component complex unit vector $Z$. To act the spinor,  $\hat{\Sigma}$ and $\hat{\gamma}^\mu$ are defined by 
\begin{align}
    \hat{\Sigma}\equiv I_{2}\otimes \Sigma,\ \hat{\gamma}^\mu\equiv \gamma^\mu\otimes I_{N+1}
\end{align}
where $\gamma^\mu$ are the gamma matrices satisfying the Clifford algebra $\qty{\gamma^\mu,\gamma^\nu}=2\eta^{\mu\nu}$ with the Minkowski metric $\eta^{\mu\nu}=\mathrm{diag}\qty(1,-1,-1)$. The two-dimensional version of the gamma matrices are defined by the usual Pauli matrices $\sigma_i~\qty(i=1,2,3)$ as $\gamma^0=\sigma_3$, $\gamma^1=-i\sigma_1$ and $\gamma^2=-i\sigma_2$ in the minimal representation. In this work, we employ the potential term $V$ of the form
\begin{equation}
    V=\frac{\mu^2}{16}\qty[\mathrm{Tr}\qty(I_{N+1}-\Sigma_\infty \Sigma)]^2
    \label{pot}
\end{equation}
where $\mu^2>0$ and $\Sigma_\infty$ is the vacuum value of $\Sigma$ at the spatial infinity. This potential is square of the so-called old-baby potential. We employ this because the ordinary old-baby potential does not support $\CP^N\ \qty(N>1)$ soliton solutions~\cite{PhysRevD.92.045007}. 

There are five parameters in the coupled Lagrangian \eqref{FullLagrangian}, $m,g,M,e$ and $\mu$. The length scale of the parameters and fields are 
\begin{align*}
	\Psi:\qty[L^{-1}]&,\ U:\qty[L^0]\\
	m:\qty[L^{-1}],\ g:\qty[L^0],\ M&:\qty[L^{-\frac{1}{2}}],\ e:\qty[L^\frac{1}{2}],\ \mu:\qty[L^{-\frac{3}{2}}].
\end{align*}
Suitably rescaling the coordinates, the fields and parameters 
\begin{align*}
	x^\mu\to\frac{e}{M}x^\mu,\ \Psi\to\frac{M}{e}\Psi,\ \mu\to \frac{M^2}{e}\mu,\ m\to\frac{M}{e}m,
\end{align*}
the rescaled Lagrangian becomes 
\begin{align}
	\mathcal{L}'
    &=\overline{\Psi}\qty(i\hat{\gamma}^\mu\partial_\mu-m\hat{\Sigma})\Psi\nonumber\\
    &+g^{-2}\qty[\frac{1}{2}\mathrm{Tr}\qty(\partial_\mu \Sigma\partial^\mu \Sigma)+\mathrm{Tr}\qty(\qty[\partial_\mu \Sigma,\partial_\nu \Sigma]^2)-V],
\end{align}
where we redefine the coupling constant $g$ as $g^{-1}\to g^{-1}\slash\sqrt{Me}$. Note that all the parameters $\qty(m,g,\mu)$ and fields $\qty(\Psi,\Sigma)$ are dimensionless. In this paper, we fix $\mu=1$ for simplicity. 

The field $\Sigma$ is invariant under the local gauge transformation $Z\to e^{i\theta}Z$ with $\theta\in\mathbb{R}$. Due to the phase redundancy and the normalization condition $Z^\dagger Z=1$, the field $Z$ takes its value on $\CP^N\simeq\mathrm{SU}\qty(N+1)\slash\qty(\mathrm{SU}\qty(N)\times\mathrm{U}\qty(1))$. In this representation, the stabilizer subgroup $\mathrm{SU}\qty(N)\times\mathrm{U}\qty(1)$ describes the isotropy subgroup of a point on the target space. The nontrivial topology of $\CP^N$ arises from the $\mathrm{U}\qty(1)$ factor in this stabilizer: the relevant homotopy relation,
\begin{align}
	\pi_2\qty(\CP^N)\simeq\pi_1\qty(\mathrm{SU}\qty(N)\times\mathrm{U}\qty(1))_{\mathrm{SU}\qty(N+1)}\simeq\pi_1\qty(\mathrm{U}\qty(1))\simeq\mathbb{Z},
\end{align}
shows that the second homotopy group of the target space $\CP^N$ is inherited from the loop structure of this $\mathrm{U}\qty(1)$ fiber. Thus there is a topological index, so-called topological charge $Q$, which is given by 
\begin{align}
	Q=-\frac{i}{2\pi}\oint_{S^1} Z^\dagger dZ.\label{Top}
\end{align}
This integer labels the Skyrmion, and corresponds to the number of fermionic zero modes. This topological charge is determined by the $\mathrm{U}\qty(1)$ factor of the stabilizer subgroup, which also governs the topology in the simpler case of $\CP^1\simeq\mathrm{SU}\qty(2)\slash\mathrm{U}\qty(1)$. 

In order to obtain the field equations, we rewrite the Lagrangian \eqref{FullLagrangian} in terms of $Z=\qty(Z_1,Z_2,\cdots,Z_N,Z_{N+1})^T$ where $T$ represents transposition of the vector. In this article, we choose the vacuum as $\Sigma_\infty=\mathrm{diag}\qty(1,1,\cdots,1,-1)$ which corresponds to $Z_\infty\propto\qty(0,0,\cdots,0,1)^T$. Then, decomposing the Lagrangian into the fermionic part $\mathcal{L}_f$ and bosonic part $\mathcal{L}_b$, i.e., $\mathcal{L}=\mathcal{L}_f+g^{-2}\mathcal{L}_b$ , one can write them in terms of $Z$ as follows: 
\begin{align}
	&\begin{aligned}
		\mathcal{L}_f
        &=\overline{\Psi}\qty[i\hat{\gamma}^\mu\partial_\mu-mI_{2}\otimes \qty(I_{N+1}-2Z\otimes Z^\dagger)]\Psi,\label{fermionL}
	\end{aligned}
	\\
    &\begin{aligned}
		\mathcal{L}_b
		&=\qty(D_\mu Z)^\dagger D^\mu Z
        +16\qty(\qty(D_\mu Z)^\dagger D_\nu Z)^2\\
        &-8\qty(\qty(D_\mu Z)^\dagger D^\mu Z)^2-8\abs{\qty(D_\mu Z)^\dagger D_\nu Z}^2\\
		&-\frac{\mu^2}{4}\qty(1-\abs{Z_{N+1}}^2)^2,\label{boseL}
	\end{aligned}
\end{align}
where $D_\mu\equiv\partial_\mu-Z^\dagger\partial_\mu Z$ is the covariant derivative. Based on the variational principle with the nonlinear constraint $Z^\dagger Z=1$, we obtain the equations of motion of the form
\begin{align}
    \qty(i\gamma^\mu_{\alpha\beta}\delta_{ab}\partial_\mu-m\delta_{\alpha\beta}\delta_{ab}+2m\delta_{\alpha\beta}Z_aZ_b^*)\Psi_{\beta,b}=0,
    \label{Fermioniceq}
\end{align}
and 
\begin{align}
	&\qty(D_\mu D^\mu-Z^\dagger D_\mu D^\mu Z)Z_a\nonumber\\
	&+32D_\mu\qty{\qty[\qty(D^\mu Z)^\dagger D^\nu Z]D_\nu Z_a}\nonumber\\
	&-32Z^\dagger D_\mu\qty{\qty[\qty(D^\mu Z)^\dagger D^\nu Z]D_\nu Z} Z_a\nonumber\\
	&-16D_\mu\qty{\qty[\qty(D_\nu Z)^\dagger D^\nu Z]D^\mu Z_a+\qty[\qty(D^\nu Z)^\dagger D^\mu Z]D_\nu Z_a}\nonumber\\
	&+16Z^\dagger D_\mu\qty{\qty[\qty(D_\nu Z)^\dagger D^\nu Z]D^\mu Z+\qty[\qty(D^\nu Z)^\dagger D^\mu Z]D_\nu Z}Z_a\nonumber\\
	&-\frac{mg^2}{2}\qty[\Psi_{\alpha,b}^*Z_b\gamma^0_{\alpha\beta}\Psi_{\beta,a}-\qty(\Psi_{\alpha,b}^*\gamma^0_{\alpha\beta}Z_bZ_c^*\Psi_{\beta,c})Z_a]\nonumber\\
	&-\frac{\mu^2}{2}\qty(1-\abs{Z_{N+1}}^2)Z_{N+1}\delta_{a,N+1}=0,\label{Bosoniceq}
\end{align}
where $\alpha,\beta=1,2$ denote the spinor indices and $a,b,c=1,2,\cdots,N+1$ represent the internal indices.

\subsection*{Rotationally Symmetric Ansatz \\and Reduced Field Equations}
In this paper, we concentrate on rotationally symmetric static $\CP^2$ Skyrmions. Because of the overall phase redundancy of the $Z$ field, such configurations can generally be parametrized as\footnote{While $\CP^2$ solitons are often constructed using local coordinates $\qty(u_1,u_2)$~\cite{Ferreira:2010jb}, we parametrize the homogeneous coordinate $Z$. When one uses the local coordinates, the boundary values at the origin and infinity involve divergence, requiring additional redefinitions for numerical computations~\cite{PhysRevD.92.045007}. To save this effort, we adopt the expression in terms of $Z$, but essentially the same thing can be done using local coordinates.}  
\begin{align}
	Z
	=\mqty
	(
		\cos F\qty(r)\\
		\sin F\qty(r)\cos G\qty(r) e^{in_1\varphi}\\
		\sin F\qty(r)\sin G\qty(r) e^{in_2\varphi}
	).
	\label{Bosonicansatz}
\end{align}
Here, we assume the profile functions $F\qty(r)$ and $G\qty(r)$ are monotonic functions of the radial coordinate $r$, $n_1$ and $n_2$ are integers, and $\varphi$ is the angular coordinate. The rotationally symmetric $\CP^2$ ansatz introduced above naturally includes $\CP^1$ configurations as a special subclass. In order to make this explicit, we describe how a $\CP^1$ embedding is realized within the present parametrization. Here we set $n_1=0$ and $G\qty(r)=\frac{\pi}{2}$, i.e., $Z_2=0$. Then we obtain a $\CP^1$ configuration $Z^{\CP^1}$ as 
\begin{align}
	Z^{\CP^1}=\mqty
	(
		\cos F\qty(r)\\
		0\\
		\sin F\qty(r)e^{in_2\varphi}
	).
\end{align}
This embedding is used as a convenient reference when comparing $\CP^1$ and $\CP^2$ configurations. 

The topological charge can be written as an integral of a total derivative. From Eq.~\eqref{Top} and Eq.~\eqref{Bosonicansatz}, we obtain
\begin{align}
    Q&=\int_0^\infty\dd{r}\partial_r\Omega,\label{Top2}\\ 
    \Omega&\equiv \sin^2F\qty(n_1\cos^2G+n_2\sin^2G).\label{Omega}
\end{align}
As a result, the topological charge $Q$ is determined by the boundary values, $\Omega\qty(0)$ and $\Omega\qty(\infty)$.

Substituting the ansatz \eqref{Bosonicansatz} into Eq.~\eqref{boseL}, we obtain the static energy density $\mathcal{E}_b=-\mathcal{L}_b$ 
\begin{align}
	\mathcal{E}_b
	&=\qty(F')^2+\sin^2F\qty(G')^2\nonumber\\
	&+\frac{1}{r^2}\qty[\frac{\Omega^2}{\tan^2F}+\qty(n_1-n_2)^2\sin^2F\sin^2G\cos^2G]\nonumber\\
	&+\frac{48}{r^2}\qty[\frac{\Omega}{\tan F}F'-\qty(n_1-n_2)\sin^2F\sin G\cos GG']^2\nonumber\\
	&+\frac{\mu^2}{4}\qty(1-\sin^2F\sin^2G)^2,\label{boseE}
\end{align}
where the derivative with respect to $r$ is denoted by $\dv{}{r}\equiv ~'$. Formally, the Euler-Lagrange equation of Eq.~\eqref{boseE} can be written as 
\begin{align}
	c_1F''+c_2G''+c_3\qty(F')^2+c_4\qty(G')^2&\nonumber\\
	+c_5F'G'+\frac{c_6}{r}F'+\frac{c_7}{r}G'+c_8&=0,\label{eqF}\\
	c_{9}G''+c_{10}F''+c_{11}\qty(G')^2+c_{12}\qty(F')^2&\nonumber\\
	+c_{13}G'F'+\frac{c_{14}}{r}G'+\frac{c_{15}}{r}F'+c_{16}&=0,\label{eqG}
\end{align}
where the coefficients $c_i~\qty(i=1,2,\cdots,16)$ are complicated trigonometric polynomials and their explicit forms are given in Appendix \ref{appendixA}. 

Since the rotationally symmetric ansatz for Skyrmion is introduced in this section, the fermion field $\Psi$ is also parametrized with the rotational symmetry. Recalling that the Skyrmions are static, we impose that the fermion field $\Psi$ has the stationary form $\Psi=\psi e^{-i\varepsilon t}$. Then, the Dirac equation \eqref{Fermioniceq} becomes 
\begin{align}
	\mathcal{H}\psi=\varepsilon\psi,\ \mathcal{H}=\hat{\gamma}^0\qty(-i\hat{\gamma}^j\partial_j+m\hat{\Sigma})\label{Fermioniceq2}
\end{align}
where $\mathcal{H}$ is the Dirac Hamiltonian and $\varepsilon$ is the energy eigenvalue. The corresponding eigenfunction of the Dirac Hamiltonian $\mathcal{H}$ can be written as
\begin{align}
	\psi=\mathcal{N}\mqty
	(
		u_1\qty(r)e^{i\ell\varphi}\\
		d_1\qty(r)e^{i\qty(\ell+n_1)\varphi}\\
		s_1\qty(r)e^{i\qty(\ell+n_2)\varphi}\\
		u_2\qty(r)e^{i\qty(\ell+1)\varphi}\\
		d_2\qty(r)e^{i\qty(\ell+n_1+1)\varphi}\\
		s_2\qty(r)e^{i\qty(\ell+n_2+1)\varphi}
	)\label{Fermionicansatz}
\end{align}
where $u_{i}$, $d_{i}$ and $s_{i}$ $\qty(i=1,2)$ are the radial functions of spinor components, $\ell\in\mathbb{Z}$ and $\mathcal{N}$ is the normalization factor which is defined by 
\begin{align}
	\int\dd[2]{x}\psi^\dagger\psi&=2\pi\mathcal{N}^2\int_{0}^{\infty}\dd{r}r\qty(u_1^2+d_1^2+s_1^2+u_2^2+d_2^2+s_2^2)\nonumber\\
	&=1.
\end{align}
Note that the Lagrangian $\mathcal{L}_f+g^{-2}\mathcal{L}_b$ on our ansatz \eqref{Bosonicansatz} and \eqref{Fermionicansatz} is invariant under a sign flip of the windings, $n_1\to-n_1$ and $n_2\to-n_2$ which corresponds to a sign reversal of the topological charge $Q$.

In this parametrization, the Dirac Hamiltonian $\mathcal{H}$ is commuted with an extended spin operator $\mathcal{K}$, so-called grand-spin operator, which is defined by 
\begin{align}
    \mathcal{K}=-i\partial_\varphi I_2\otimes I_3+\frac{1}{2}\sigma_3\otimes I_3+\frac{n_1}{2}I_2\otimes\lambda_3-\frac{n_1-2n_2}{2\sqrt{3}}I_2\otimes\lambda_8 
\end{align}
where $\lambda_3$ and $\lambda_8$ represent the third and eighth components of Gell-Mann matrices respectively. Since this operator satisfies $\mathcal{K}\mathcal{H}\mathcal{K}^{-1}=\mathcal{H}$, the solutions of Eq.~\eqref{Fermioniceq2} are labeled by the eigenvalue $\kappa$ which is given by
\begin{align}
	\mathcal{K}\psi&=\kappa\psi,\\
    \kappa&=\ell+\frac{1}{2}+\frac{n_1+n_2}{3}.
\end{align}

Plugging the ansatz \eqref{Bosonicansatz} and \eqref{Fermionicansatz} into the equations of the system \eqref{Bosoniceq} and \eqref{Fermioniceq2}, we can obtain some ODEs 

\begin{widetext}
\begin{align}
	&c_1F''+c_2G''+c_3\qty(F')^2+c_4\qty(G')^2+c_5F'G'+\frac{c_6}{r}F'+\frac{c_7}{r}G'+c_8\nonumber\\
	&\qquad-\frac{mg^2}{2}r\qty[-\qty(u_1^2-u_2^2)\sin2F+\qty(u_1d_1-u_2d_2)\cos2F\cos G+\qty(u_1s_1-u_2s_2)\cos2F\sin G]\nonumber\\
	&\qquad-\frac{mg^2}{2}r\qty[\qty(u_1d_1-u_2d_2)\cos2F\cos G+\qty(d_1^2-d_2^2)\sin2F\cos^2G+\qty(d_1s_1-d_2s_2)\sin2F\sin G\cos G]\nonumber\\
	&\qquad-\frac{mg^2}{2}r\qty[\qty(u_1s_1-u_2s_2)\cos2F\sin G+\qty(d_1s_1-d_2s_2)\sin2F\sin G\cos G+\qty(s_1^2-s_2^2)\sin2F\sin^2G]
	=0,\label{coupledODE1}\\
	&c_9G''+c_{10}F''+c_{11}\qty(G')^2+c_{12}\qty(F')^2+c_{13}G'F'+\frac{c_{14}}{r}G'+\frac{c_{15}}{r}F'+c_{16}\nonumber\\
	&\qquad-\frac{mg^2}{2}r\qty[-\qty(u_1d_1-u_2d_2)\sin F\cos F\sin G+\qty(u_1s_1-u_2s_2)\sin F\cos F\cos G]\nonumber\\
	&\qquad-\frac{mg^2}{2}r\qty[-\qty(u_1d_1-u_2d_2)\sin F\cos F\sin G-\qty(d_1^2-d_2^2)\sin^2F\sin2G+\qty(d_1s_1-d_2s_2)\sin^2F\cos2G]\nonumber\\
	&\qquad-\frac{mg^2}{2}r\qty[\qty(u_1s_1-u_2s_2)\sin F\cos F\cos G+\qty(d_1s_1-d_2s_2)\sin^2F\cos2G+\qty(s_1^2-s_2^2)\sin^2F\sin2G]=0,\label{coupledODE2}
\end{align}
\begin{align}
	&m\qty(1-2\cos^2F)u_1+m\qty(-2\sin F\cos F\cos G)d_1+m\qty(-2\sin F\cos F\sin G)s_1
	+\qty(-\partial_r-\frac{\ell+1}{r})u_2
	=\varepsilon u_1,\label{coupledODE3}\\
	&m\qty(-2\sin F\cos F\cos G)u_1+m\qty(1-2\sin^2 F\cos^2 G)d_1+m\qty(-2\sin^2F\sin G\cos G)s_1
	+\qty(-\partial_r-\frac{\ell+n_1+1}{r})d_2
	=\varepsilon d_1,\label{coupledODE4}\\
	&m\qty(-2\sin F\cos F\sin G)u_1+m\qty(-2\sin^2F\sin G\cos G)d_1+m\qty(1-2\sin^2F\sin^2G)s_1
	+\qty(-\partial_r-\frac{\ell+n_2+1}{r})s_2
	=\varepsilon s_1,\label{coupledODE5}\\
	&m\qty(1-2\cos^2F)u_2+m\qty(-2\sin F\cos F\cos G)d_2+m\qty(-2\sin F\cos F\sin G)s_2
	+\qty(-\partial_r+\frac{\ell}{r})u_1
	=-\varepsilon u_2,\label{coupledODE6}\\
	&m\qty(-2\sin F\cos F\cos G)u_2+m\qty(1-2\sin^2 F\cos^2 G)d_2+m\qty(-2\sin^2F\sin G\cos G)s_2
	+\qty(-\partial_r+\frac{\ell+n_1}{r})d_1
	=-\varepsilon d_2,\label{coupledODE7}\\
	&m\qty(-2\sin F\cos F\sin G)u_2+m\qty(-2\sin^2F\sin G\cos G)d_2+m\qty(1-2\sin^2F\sin^2G)s_2
	+\qty(-\partial_r+\frac{\ell+n_2}{r})s_1
	=-\varepsilon s_2.\label{coupledODE8}
\end{align}
\end{widetext}
Solutions of these equations give the stationary point of the total energy functional 
\begin{align}
    E_{\mathrm{tot}}=\int\dd[2]{x}\qty(\psi^\dagger\mathcal{H}\psi+g^{-2}\mathcal{E}_b).
    \label{eq:total_energy}
\end{align}
In order to obtain the solutions numerically, some appropriate boundary conditions, at the origin and the spatial infinity, are required. 

\section{Numerical Scheme}
\label{NumericalMethod}
In this section, we describe the numerical scheme we used to solve the coupled ODEs \eqref{coupledODE1}-\eqref{coupledODE8} self-consistently. The subsection~\ref{subsecA} and \ref{subsecB} explain the boundary conditions for the Skyrmion field and fermion field respectively. The subsection~\ref{subsecC} illustrates the numerical method. 

\subsection{Boundary Conditions for Profile Functions}
\label{subsecA}
As we wrote above, we impose $Z_\infty\propto\qty(0,0,1)^T$. It implies that $F$ and $G$ must satisfy 
\begin{align}
	F\qty(\infty)=\qty(\frac{1}{2}+p)\pi,\ G\qty(\infty)=\qty(\frac{1}{2}+q)\pi\ \qty(p,q\in\mathbb{Z}).
\end{align}
The $\CP^2$ Skyrmions can be constructed with the following boundary conditions at the origin: 
\begin{align}
	F(0)=p\pi,\ G(0)=q\pi\ \qty(p,q\in\mathbb{Z})\label{BSFG0}
\end{align}
For simplicity, we set $p=q=0$, and therefore the boundary conditions we impose for the $\CP^2$ Skyrmions are 
\begin{align}
	F\qty(0)=G\qty(0)=0,\ F\qty(\infty)=G\qty(\infty)=\frac{\pi}{2}.
\end{align}
Then, it follows from Eqs.~\eqref{Top2} and \eqref{Omega} that the topological charge is 
\begin{align}
    Q&=\Omega\qty(\infty)-\Omega\qty(0)=n_2.\label{Qtop}
\end{align}
In this case, $n_1$ is not a topological quantity. Note that for the $\CP^1$ embedding configuration, we set $G\qty(r)=\pi/2$, instead of $G(0)=0$. One can see that the topological charge of such $\CP^1$ Skyrmions is also given by $Q=n_2$. 

\subsection{Boundary Conditions for Spinors}
\label{subsecB}
Here, we perform the asymptotic analysis for the Dirac equation. Taking into account the boundary condition $F\qty(0)$ and $G\qty(0)$ given in Eq.~\eqref{BSFG0}, the Dirac equation at the vicinity of origin becomes 
\begin{align}
	\qty(m+\varepsilon)u_1+\qty(\partial_r+\frac{\ell+1}{r})u_2
	&=0,\\
	\qty(-m+\varepsilon)d_1+\qty(\partial_r+\frac{\ell+n_1+1}{r})d_2
	&=0,\\
	\qty(-m+\varepsilon)s_1+\qty(\partial_r+\frac{\ell+n_2+1}{r})s_2
	&=0,\\
	\qty(m-\varepsilon)u_2+\qty(\partial_r-\frac{\ell}{r})u_1
	&=0,\\
	\qty(-m-\varepsilon)d_2+\qty(\partial_r-\frac{\ell+n_1}{r})d_1
	&=0,\\
	\qty(-m-\varepsilon)s_2+\qty(\partial_r-\frac{\ell+n_2}{r})s_1
	&=0.
\end{align}
Assuming that each spinor component can be expanded around the origin in a power series, we can find relations between the boundary values of the spinor components and the angular momentum $\ell$ of the form
\begin{align}
	\ell u_1\qty(0)=0,\ \qty(\ell+1) u_2\qty(0)=0,\\
	\qty(\ell+n_1) d_1\qty(0)=0,\ \qty(\ell+n_1+1) d_2\qty(0)=0,\\
	\qty(\ell+n_2) s_1\qty(0)=0,\ \qty(\ell+n_2+1) s_2\qty(0)=0.
\end{align}

For the embedded $\CP^1$ case, we set $G\qty(r)=\frac{\pi}{2}$ and $n_1=0$. Then we have the Dirac equations
\begin{align}
	m\qty(1-2\cos^2F)u_1+m\qty(-2\sin F\cos F)s_1&\nonumber\\
	+\qty(-\partial_r-\frac{\ell+1}{r})u_2
	&=\varepsilon u_1,\\
	md_1
	+\qty(-\partial_r-\frac{\ell+1}{r})d_2
	&=\varepsilon d_1,\\
	m\qty(-2\sin F\cos F)u_1+m\qty(1-2\sin^2F)s_1&\nonumber\\
	+\qty(-\partial_r-\frac{\ell+n_2+1}{r})s_2
	&=\varepsilon s_1,\\
	m\qty(1-2\cos^2F)u_2+m\qty(-2\sin F\cos F)s_2&\nonumber\\
	+\qty(-\partial_r+\frac{\ell}{r})u_1
	&=-\varepsilon u_2,\\
	md_2
	+\qty(-\partial_r+\frac{\ell}{r})d_1
	&=-\varepsilon d_2,\\
	m\qty(-2\sin F\cos F)u_2+m\qty(1-2\sin^2F)s_2&\nonumber\\
	+\qty(-\partial_r+\frac{\ell+n_2}{r})s_1
	&=-\varepsilon s_2.
\end{align}
These equations show that $d_1$ and $d_2$ represent plane wave solutions in this case. Since these free components may destroy the numerical calculation, because the corresponding bosonic $Z_2$ component vanishes, we set $d_1\qty(r)=d_2\qty(r)=0$. Based on this argument, we can get similar boundary conditions :
\begin{align}
	\ell u_1\qty(0)=0,\ \qty(\ell+1) u_2\qty(0)=0,\\
	\qty(\ell+n_2) s_1\qty(0)=0,\ \qty(\ell+n_2+1) s_2\qty(0)=0.
\end{align}

At the spatial infinity, the Dirac equations approach 
\begin{align}
	\qty(-m+\varepsilon)u_1+\qty(\partial_r+\frac{\ell+1}{r})u_2
	&=0,\\
	\qty(-m+\varepsilon)d_1+\qty(\partial_r+\frac{\ell+n_1+1}{r})d_2
	&=0,\\
	\qty(m+\varepsilon)s_1+\qty(\partial_r+\frac{\ell+n_2+1}{r})s_2
	&=0,\\
	\qty(-m-\varepsilon)u_2+\qty(\partial_r-\frac{\ell}{r})u_1
	&=0,\\
	\qty(-m-\varepsilon)d_2+\qty(\partial_r-\frac{\ell+n_1}{r})d_1
	&=0,\\
	\qty(m-\varepsilon)s_2+\qty(\partial_r-\frac{\ell+n_2}{r})s_1
	&=0.
\end{align}
These equations have analytical solutions written by the standard Bessel function. Using these solutions and the appropriate boundary condition for the Bessel function, one can construct the spinor basis in a cylinder of radius $R$ as shown in Refs.~\cite{Amari:2019tgs,Kahana:1984be,Kahana:1984dx,Amari:2023gjq,Amari:2024rpm}. The explicit form of the plane wave basis is presented in Appendix~\ref{appendixB}. 

\subsection{Numerical Methods}
\label{subsecC}
Here, we provide a quick overview of the numerical procedure we use to solve the coupled ODEs \eqref{coupledODE1}-\eqref{coupledODE8}. Instead of solving the Dirac equation~\eqref{coupledODE3}-\eqref{coupledODE6} directly, we solve the following secular equation, which corresponds to the eigenequation \eqref{Fermioniceq2}. The secular equation for the energy eigenvalue $\varepsilon$ is given by
\begin{align}
	\det\qty(H-\varepsilon \Tilde{I})=0,\label{Secular}
\end{align}
where the matrix element of $H$ and $\Tilde{I}$ is defined as 
\begin{align}
	H_{k'^{\qty(p)},k^{\qty(q)}}&\equiv\int\dd[2]{x}\Phi_{\kappa}^{\qty(p)}\qty(k'^{\qty(p)},x)^\dagger\mathcal{H}\Phi_{\kappa}^{\qty(q)}\qty(k^{\qty(q)},x),\\
	\Tilde{I}_{k'^{\qty(p)},k^{\qty(q)}}&\equiv\int\dd[2]{x}\Phi_{\kappa}^{\qty(p)}\qty(k'^{\qty(p)},x)^\dagger\Phi_{\kappa}^{\qty(q)}\qty(k^{\qty(q)},x),\\ 
	&\qty(p,q=u,d,s),\nonumber
\end{align}
where $\Phi_{\kappa}^{\qty(q)}$ is the orthonormal basis with the quantum number $\kappa$ and $k^{\qty(q)}$ is discretized wavenumber associated with the boundary condition of the Bessel function (For details, see Ref.~~\cite{Amari:2019tgs,Kahana:1984be,Kahana:1984dx,Amari:2023gjq,Amari:2024rpm} and Appendix~\ref{appendixB}.). 

By solving the secular equation~\eqref{Secular} numerically, we can obtain some energy spectra $\qty{\varepsilon}$ and corresponding eigenspinors $\qty{\psi}$ for each $\kappa$. The eigenproblem~\eqref{Secular} is solved by a standard matrix diagonalization solver 'dsyev' in the LAPACK, Math Kernel Library. This simulation uses the computational domain-size $R=25.0$, and the number of discrete wavenumbers is chosen as $512$, which gives sufficient convergence.

In order to solve the ODEs \eqref{coupledODE1} and \eqref{coupledODE2} for the Skyrmion field, we use the Newton-Raphson method with $512$ mesh points. The computational errors is of order $10^{-6}$. Since the Skyrmion localizes around the origin, we impose the boundary condition $F\qty(R)=G\qty(R)=\frac{\pi}{2}$ at $r=R$ instead of the spatial infinity. 

We solve the coupled equations \eqref{coupledODE1}-\eqref{coupledODE8} by combining the above two numerical calculation methods. The numerical procedure consists of the following steps. 
\begin{enumerate}
	\item Prepare a suitable initial configuration $F_{\mathrm{ini}}$ and $G_{\mathrm{ini}}$.
    \footnote{We choose an analytical solution of the nonlinear sigma model $\mathcal{L}=\frac{M^2}{2}\mathrm{Tr}\qty(\partial_\mu\Sigma\partial^\mu\Sigma)$ as the initial condition~\cite{Polyakov:1975yp,Golo:1978de,DAdda:1978vbw,Perelomov:1987va,Rajaraman:1982is}.}
	\item  Solve the ODEs for $F$ and $G$ without backreaction \eqref{eqF} and \eqref{eqG} by the Newton-Raphson method with the initial configurations $F_{\mathrm{ini}}$ and $G_{\mathrm{ini}}$. 
	\item  Solve the secular equation \eqref{Secular} with the matrix diagonalization method, using the Skyrmion configuration obtained in Step 2, as the background field.
	\item Solve the Skyrmion field equations \eqref{coupledODE1} and \eqref{coupledODE2} by the Newton-Raphson method using the fermion configuration $\qty{\psi_\kappa,\varepsilon_\kappa}$ with specific $\kappa$ obtained in Step 3 as the background field.
	\item  Solve the secular equation \eqref{Secular} with the matrix diagonalization method, using the Skyrmion configuration obtained in Step 4 as the background field.
	\item Calculate the relative errors between the fermion configurations obtained in Step $3$ and Step $5$. If the relative error is sufficiently small (we employ a criterion of order $10^{-6}$), 
    the configuration can be regarded as a solution in a self-consistent manner.

    Otherwise, repeat Steps $4$ and $5$ until the relative errors between 
	fermion configurations in successive iterations satisfy the convergence criterion. 
\end{enumerate}
Note that, in this paper, we consider the coupled system with a single fermion mode. One can generalize this analysis toward more general situations that multi fermion modes couple to the Skyrmion. The detail is explained in the next section. 
\begin{figure}[t]
	\centering
		\includegraphics[width=1\linewidth]{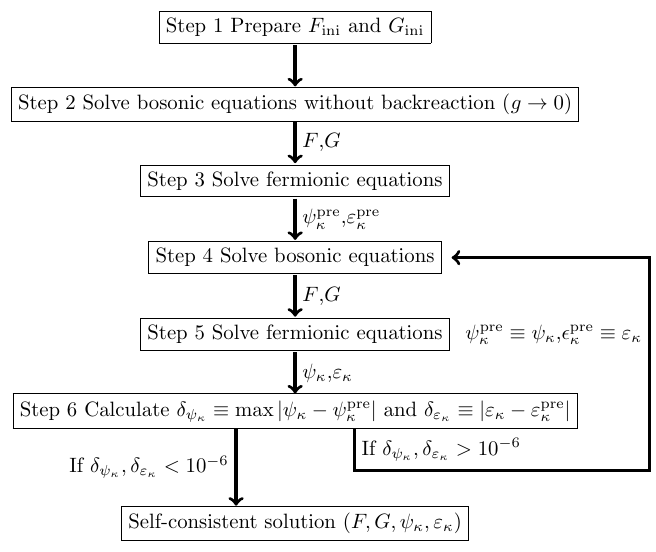}
		\caption{Schematic illustration of the self-consistent analysis}
\end{figure}



\section{Numerical Results}
\label{main}

Here we present our typical numerical solutions. 
Throughout this paper, we consider six combinations $\qty(n_1,n_2)$ as shown in Table~\ref{Table_n1n2}, where $n_2$ is equal to the topological charge. The cases of $n_1=0$ corresponds to the embedded $\CP^1$ cases with $G\qty(r)=\pi/2$. On the other hand, the genuine $\CP^2$ Skyrmions possess $n_1\neq 0$. In this paper, we consider two genuine $\CP^2$ cases with $\qty(n_1,n_2)=\qty(\pm1,\pm3)$, which gives the simplest genuine $\CP^2$ Skyrmions in our model\footnote{Although static bosonic solutions with $\abs{Q}=2$ can formally be constructed, it has been shown that such configurations exhibit pathological behavior once time-dependence is considered, leading to divergent energies when interpreted as vortices~\cite{PhysRevD.92.045007}.}. In order to compare the $\CP^1$ cases with the $\CP^2$ cases, we also consider the simplest embedding cases, $\qty(n_1,n_2)=\qty(0,\pm1)$, and the $\CP^1$ Skyrmions possessing the same topological charge with the genuine $\CP^2$ Skyrmions, i.e., $\qty(n_1,n_2)=\qty(0,\pm 3)$. 

\begin{figure}[t]
    \centering
    \includegraphics[width=1\linewidth]{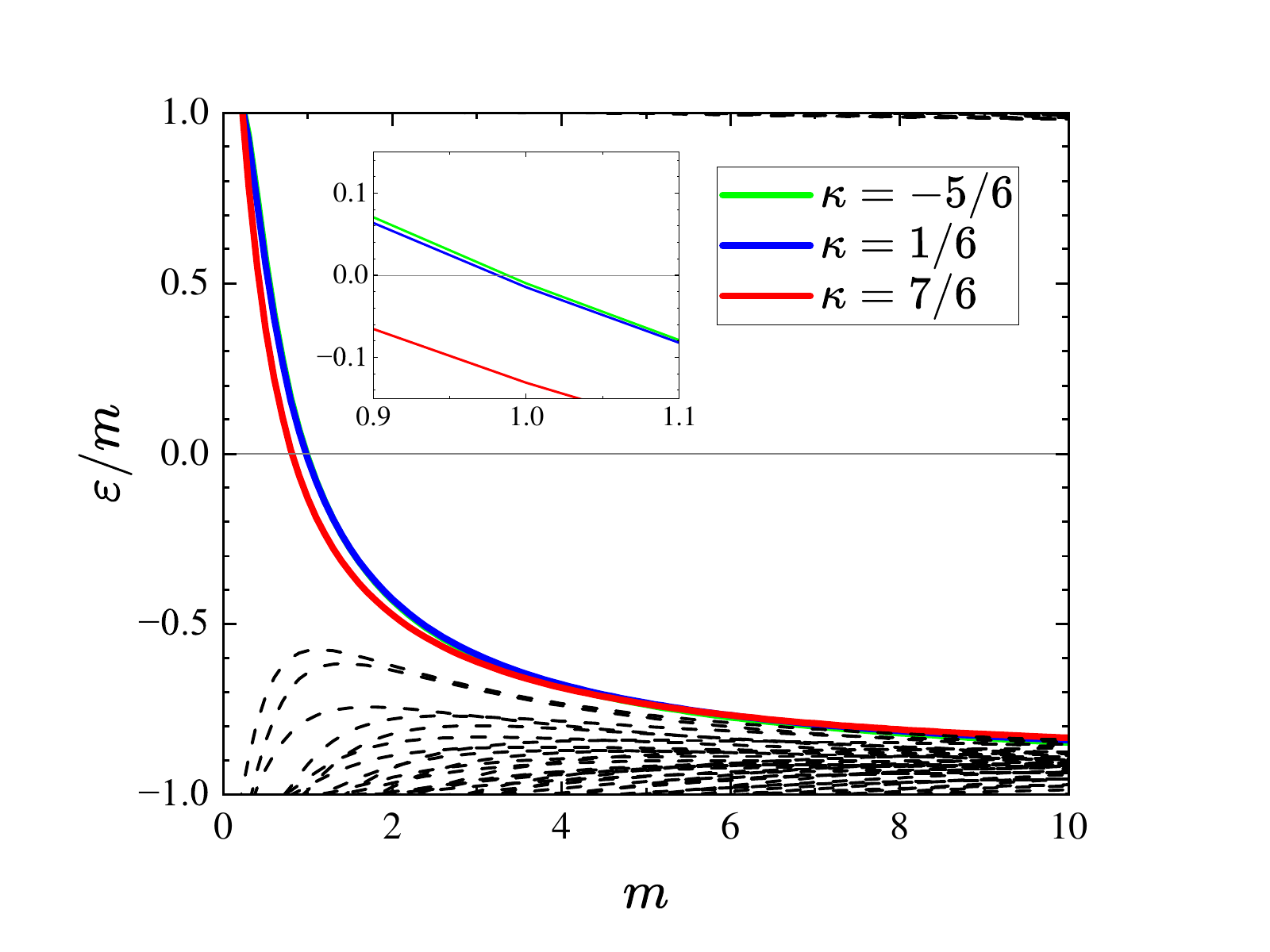}
    \caption{An example of the spectral flow on a background Skyrmion with $\qty(n_1,n_2)=\qty(-1,-3)$.}
    \label{flow_background_n1-1_n2-3}
\end{figure}

\begin{table}[H]
	\centering
	\caption{\centering The sets of the winding numbers $\qty(n_1,n_2)$ that we consider in this paper. The integer $n_2$ equals to the topological charge.}
	\label{Table_n1n2}
		\begin{tabular}{c|cccccc}
			\hline\hline
			$n_1$&~$0$~&~$0$~&~$0$~&~$0$~&~$1$~&~$-1$~\\
			\hline
			$n_2$&~$1$~&~$-1$~&~$3$~&~$-3$~&~$3$~&~$-3$~\\
			\hline\hline
		\end{tabular}
\end{table}
As for the fermion solutions, we define the fermion density $\rho_\kappa$ which is given by 
\begin{align}
	\rho_\kappa\qty(r)
	&=\frac{1}{2\pi}\int_0^\infty\dd{\varphi}\psi_\kappa^\dagger\psi_\kappa.
\end{align}
Thanks to the index theorem, the fermion energy spectra $\qty{\varepsilon}$ possess a number of zero-crossing modes corresponding to the topological charge of Skyrmions as shown in Fig.~\ref{flow_background_n1-1_n2-3} (see also Ref.~\cite{Amari:2019tgs,Amari:2023gjq,Amari:2024rpm}). Since our primary interest is to clarify the qualitative difference between cases with positive and negative topological charges, we extract the essence of the asymmetry by choosing a single zero mode, i.e., a configuration in which only a single fermionic zero mode is occupied. The choices of quantum number $\kappa$ for the variety of $\qty(n_1,n_2)$ are shown in Table.~\ref{Table_kappa}. Note that the $\abs{Q}$ smallest  values of $\abs{\kappa}$ correspond to the zero modes. For $\abs{Q}>1$, the Skyrmion is no longer concentrated at the origin but develops a ring-shaped profile peaked at a finite radius~\cite{Piette:1994,Manton:2004tk}. To align the fermion localization with this structure, it is natural to couple the Skyrmion to a zero mode whose radial profile also peaks away from the origin. Since the zero modes with larger quantum number $\abs{\kappa}$ localize at larger radii, we choose the mode with the highest $\abs{\kappa}$, which achieves the best spatial overlap with the ring-like Skyrmion core and therefore maximizes the backreaction.
\begin{table}[H]
	\centering
	\caption{\centering The choices of the grand spin quantum number $\kappa$ for each combinations of $(n_1,n_2)$}
	\label{Table_kappa}
		\begin{tabular}{c|cccccc}
			\hline\hline
			$\qty(n_1,n_2)$&~$\qty(0,1)$~&~$\qty(0,-1)$~&~$\qty(0,3)$~&~$\qty(0,-3)$~&~$\qty(1,3)$~&~$\qty(-1,-3)$~\\
			\hline
			$\kappa$&$-1\slash 6$&$1\slash 6$&$-3\slash 2$&$3\slash 2$&-$7\slash 6$&$7\slash 6$\\
			\hline\hline
		\end{tabular}
\end{table}

Moreover, from the spectral flow in the background approximation (see Fig.~\ref{flow_background_n1-1_n2-3}), one expects that different zero modes produce qualitatively similar backreaction effects. We have explicitly verified that this expectation holds: occupying different single zero mode leads to essentially the same deformation of the Skyrmion configuration. If all zero modes are occupied, their contributions simply add up in the coupled equations \eqref{coupledODE1} and \eqref{coupledODE2} through interaction terms of the same structure, without inducing any additional interactions among different zero modes. As a result, the fermionic backreaction differs only by an overall multiplicative factor proportional to the number of occupied modes. For this reason, restricting to a single occupied zero mode is sufficient to capture the qualitative features of the fermion-induced deformation. 

\begin{widetext}
	
	\begin{figure}[H]
		\begin{minipage}[b]{0.5\linewidth}
			\centering
			\includegraphics[width=1\linewidth]{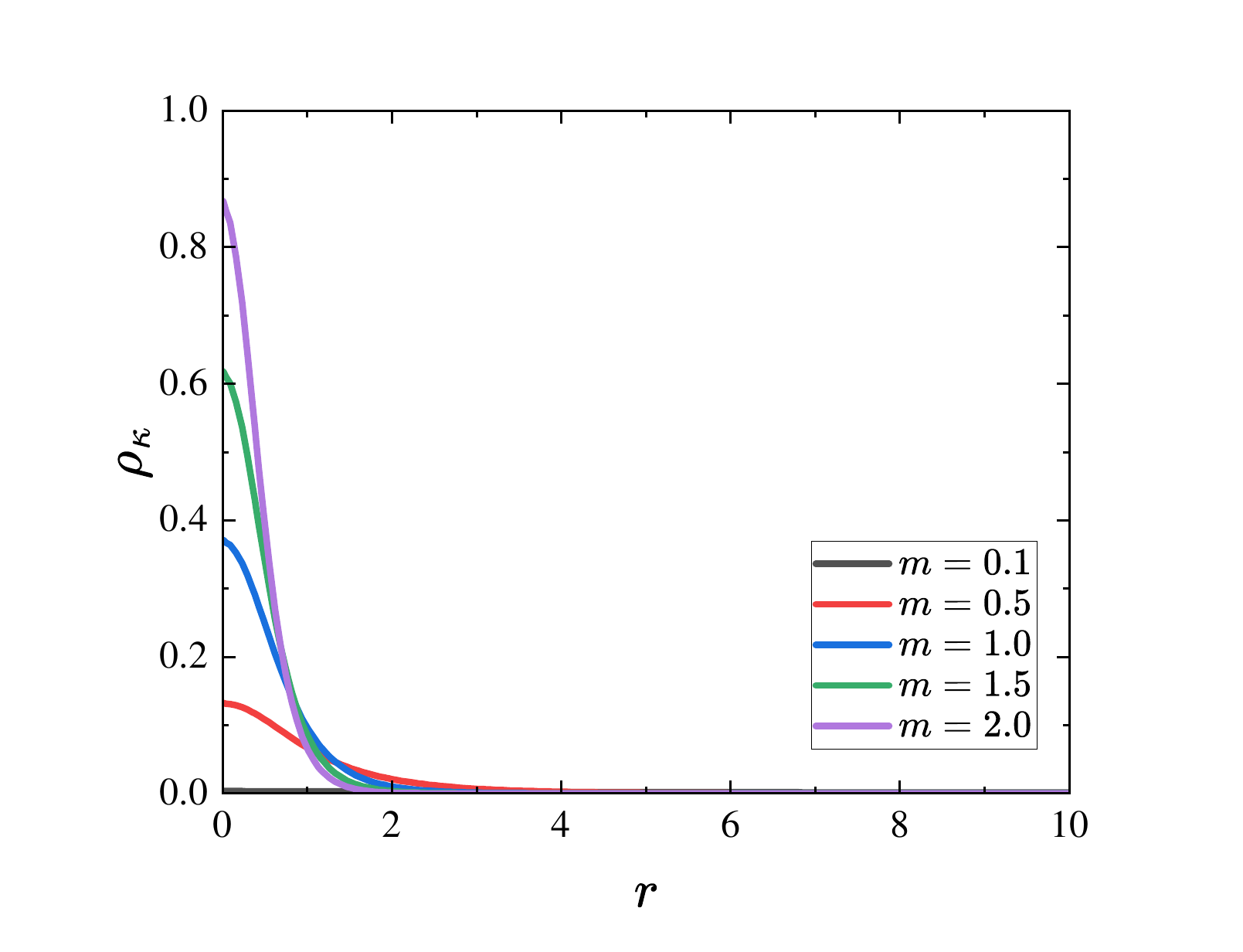}
			\subcaption{$\CP^1$ case with $Q=1\ \qty(n_1=0,n_2=1)$}
			\label{FDens_CP1_Q1_g03}
		\end{minipage}
		\begin{minipage}[b]{0.5\linewidth}
			\centering
			\includegraphics[width=1\linewidth]{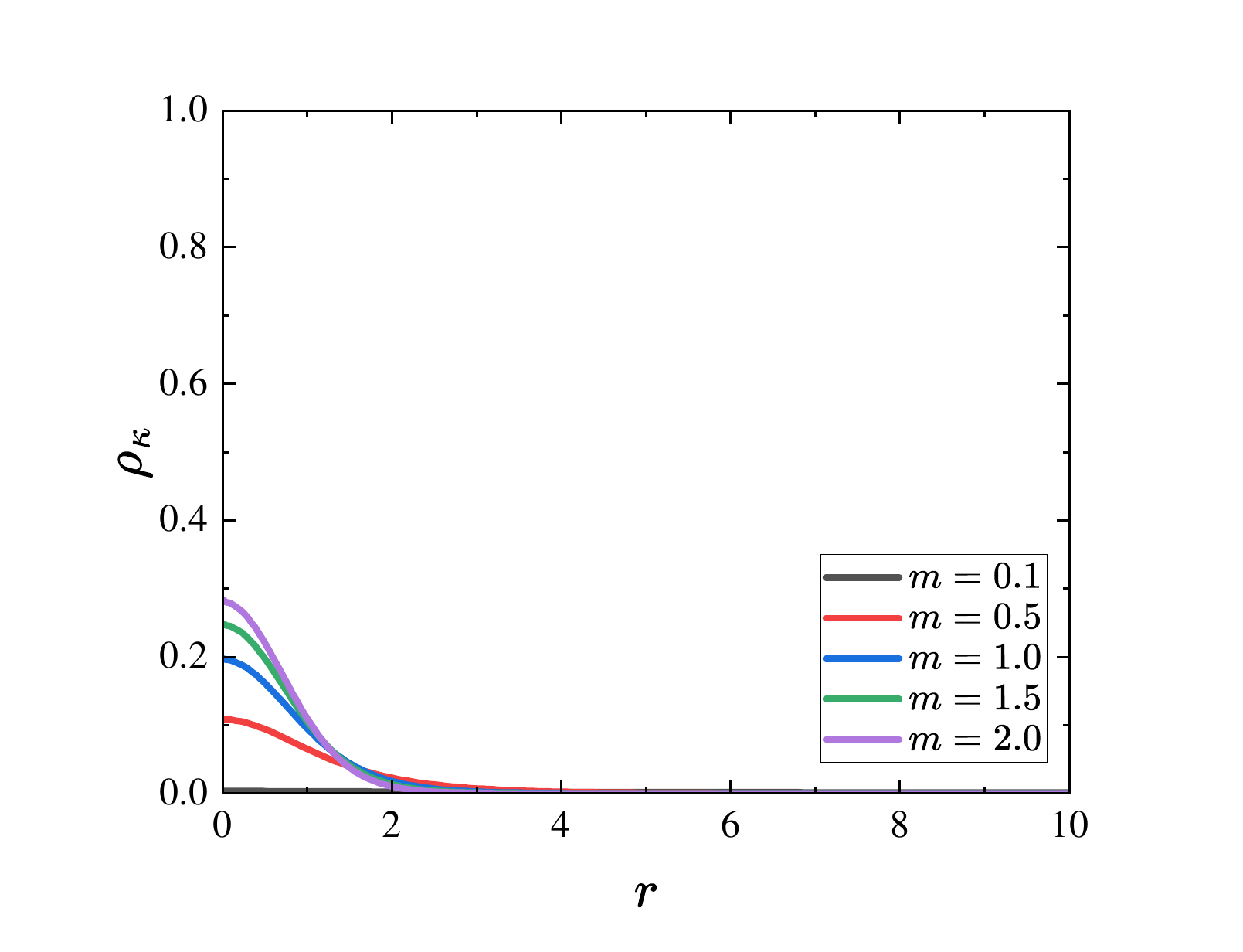}
			\subcaption{$\CP^1$ case with $Q=-1\ \qty(n_1=0,n_2=-1)$}
			\label{FDens_CP1_Q-1_g03}
		\end{minipage}\\
		\begin{minipage}[b]{0.5\linewidth}
			\centering
			\includegraphics[width=1\linewidth]{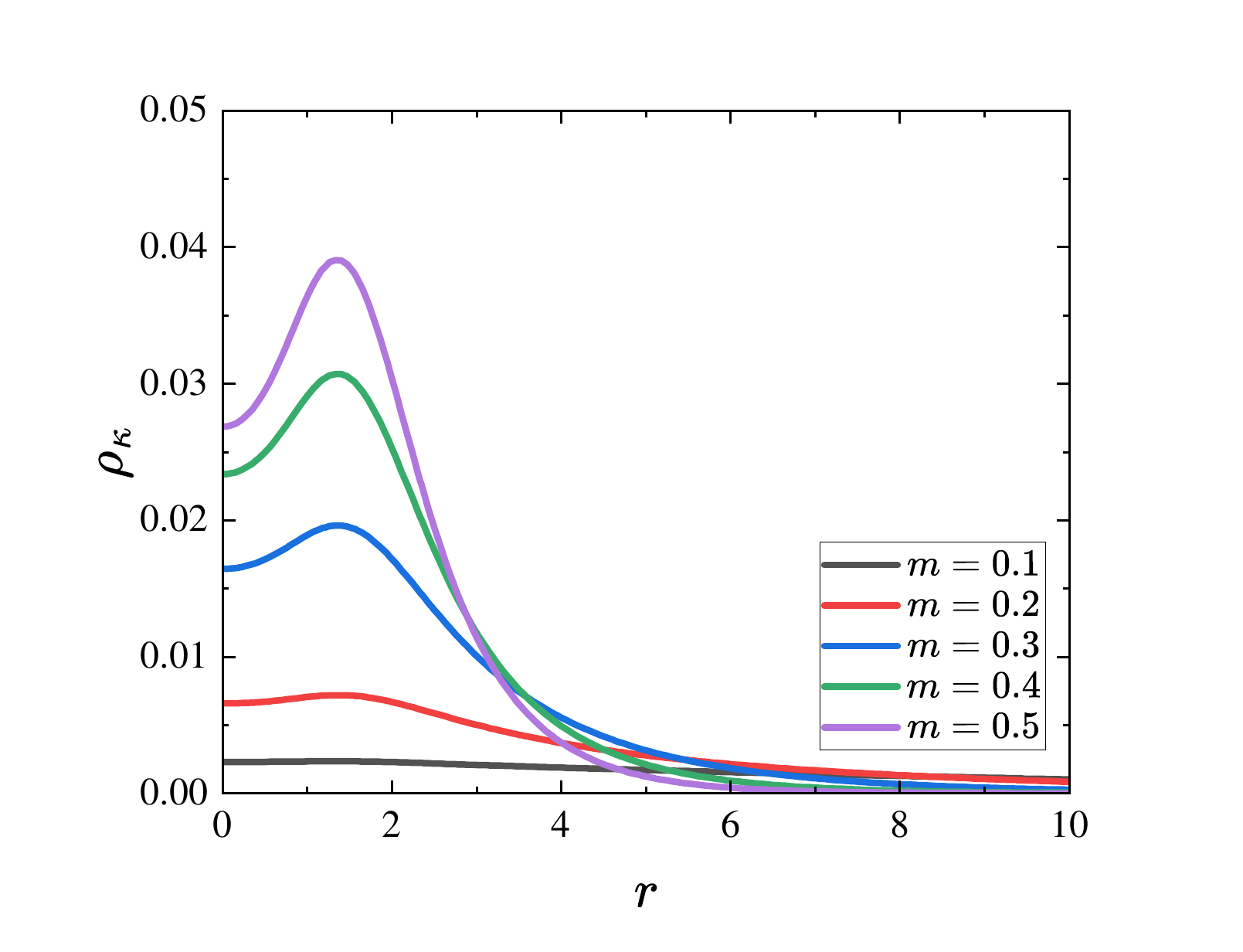}
			\subcaption{$\CP^1$ case with $Q=3\ \qty(n_1=0,n_2=3)$}
			\label{FDens_CP1_Q3_g03}
		\end{minipage}
		\begin{minipage}[b]{0.5\linewidth}
			\centering
			\includegraphics[width=1\linewidth]{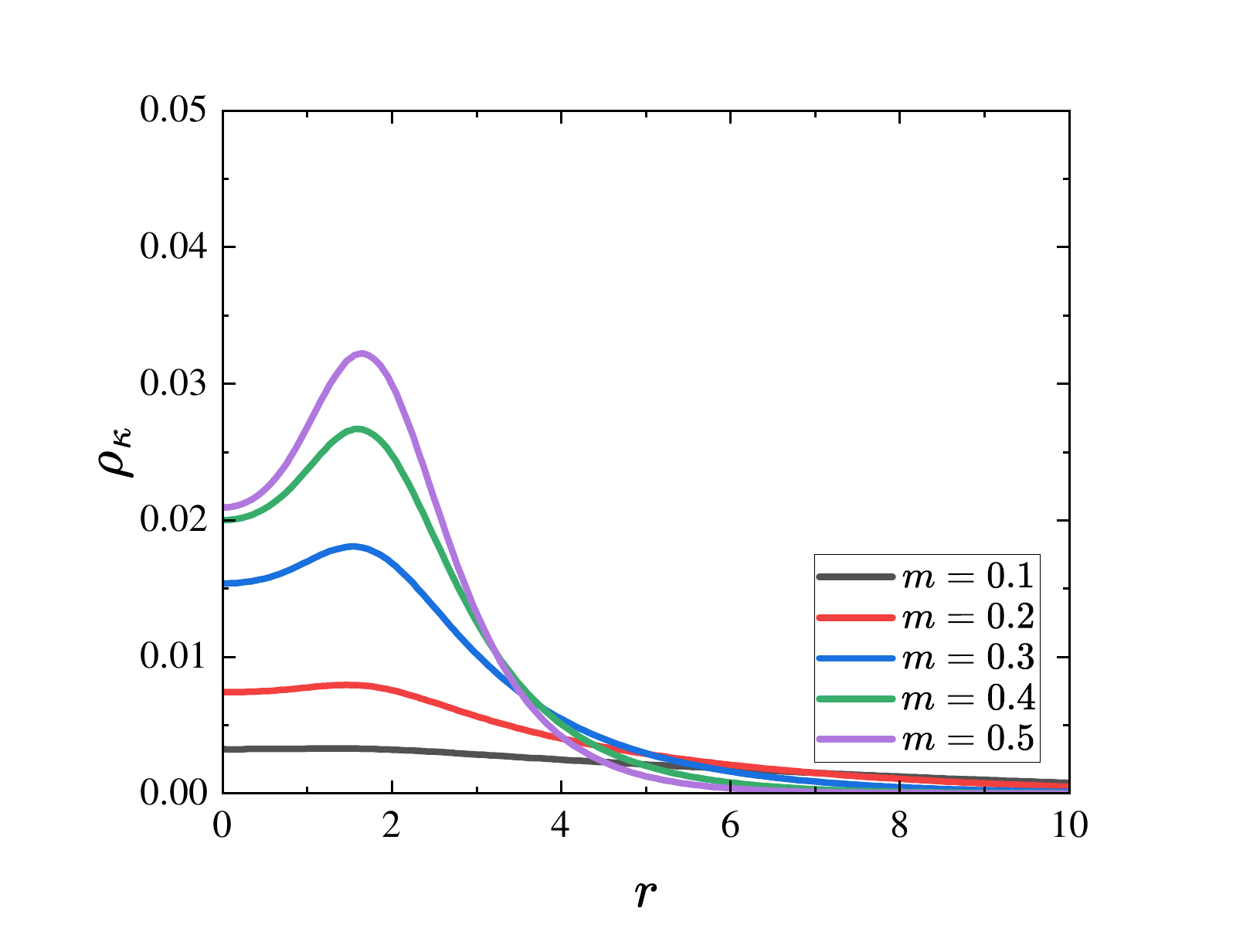}
			\subcaption{$\CP^1$ case with $Q=-3\ \qty(n_1=0,n_2=-3)$}
			\label{FDens_CP1_Q-3_g03}
		\end{minipage}\\
		\begin{minipage}[b]{0.5\linewidth}
			\centering
			\includegraphics[width=1\linewidth]{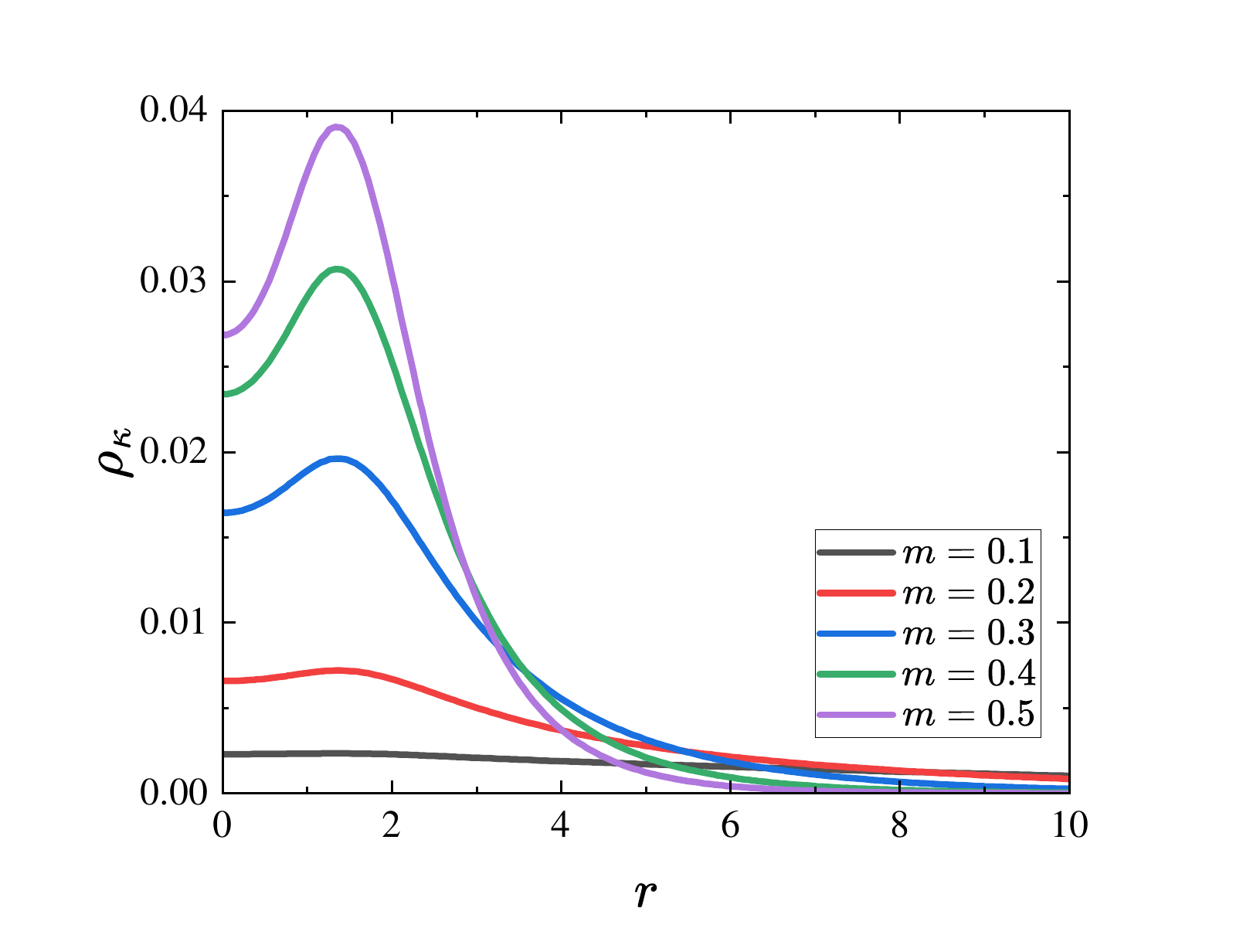}
			\subcaption{$\CP^2$ case with $Q=3\ \qty(n_1=1,n_2=3)$}
			\label{FDens_CP2_Q3_g03}
		\end{minipage}
		\begin{minipage}[b]{0.5\linewidth}
			\centering
			\includegraphics[width=1\linewidth]{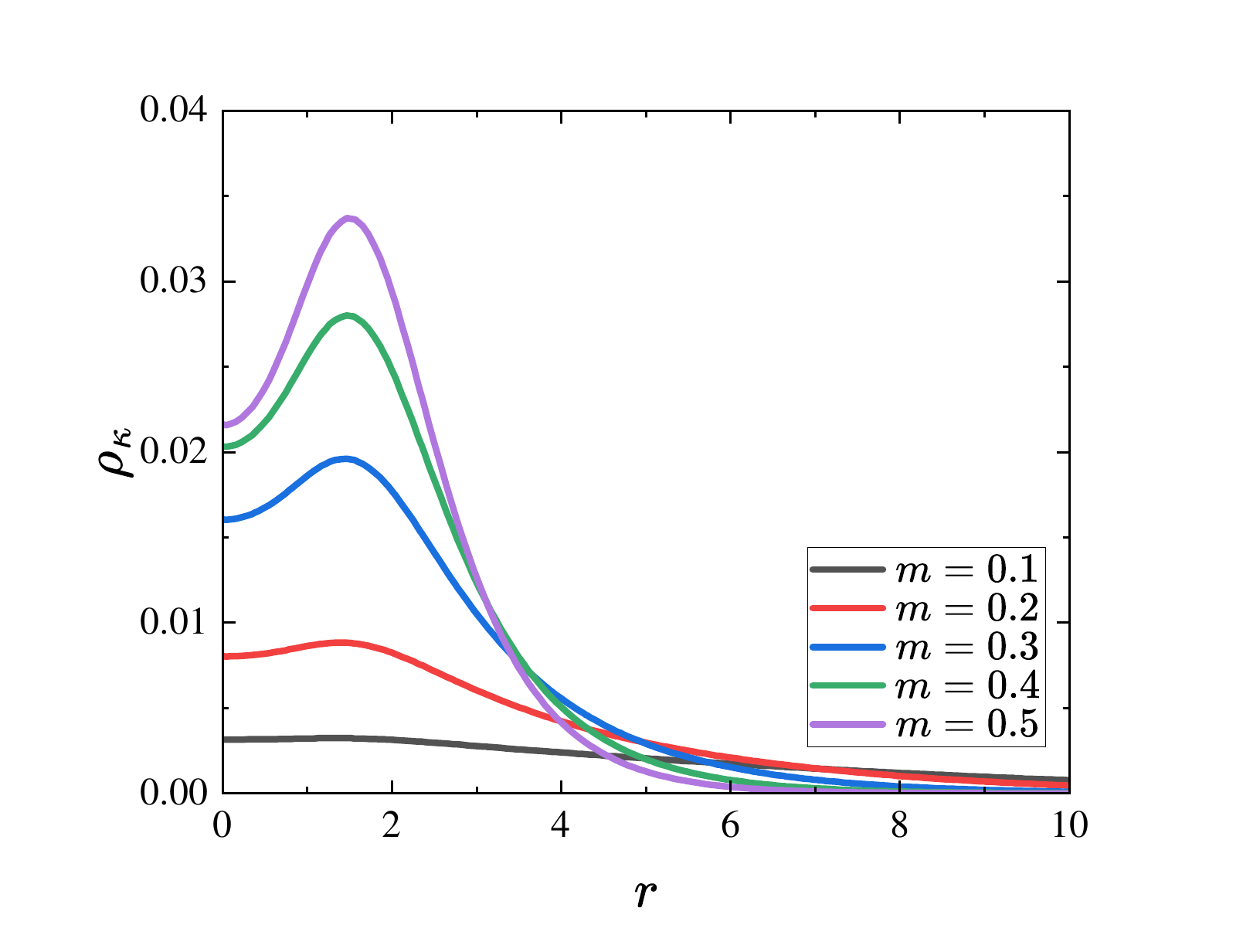}
			\subcaption{$\CP^2$ case with $Q=-3\ \qty(n_1=-1,n_2=-3)$}
			\label{FDens_CP2_Q-3_g03}
		\end{minipage}\\
        \caption{Typical fermion density $\rho_\kappa$ for several values of the Yukawa coupling constant $m$, with $g^{-2}=0.9$. For each sector, results for the same set of coupling constant $m$ are shown to allow a direct comparison between $Q>0$ and $Q<0$. } 
		\label{FDens_g03}
	\end{figure}
\end{widetext}

Our results in Fig.~\ref{FDens_g03} show that increasing the Yukawa coupling constant $m$, the fermion densities $\rho_\kappa$ begin to localize. In the weak coupling regime, $\rho_\kappa$ spreads out the space like plane waves. In the strong coupling regime, however, $\rho_\kappa$ is strongly localized around the origin. This is a consequence of the index theorem.

\begin{widetext}
	
	\begin{figure}[H]
		\begin{minipage}[b]{0.5\linewidth}
			\centering
			\includegraphics[width=1\linewidth]{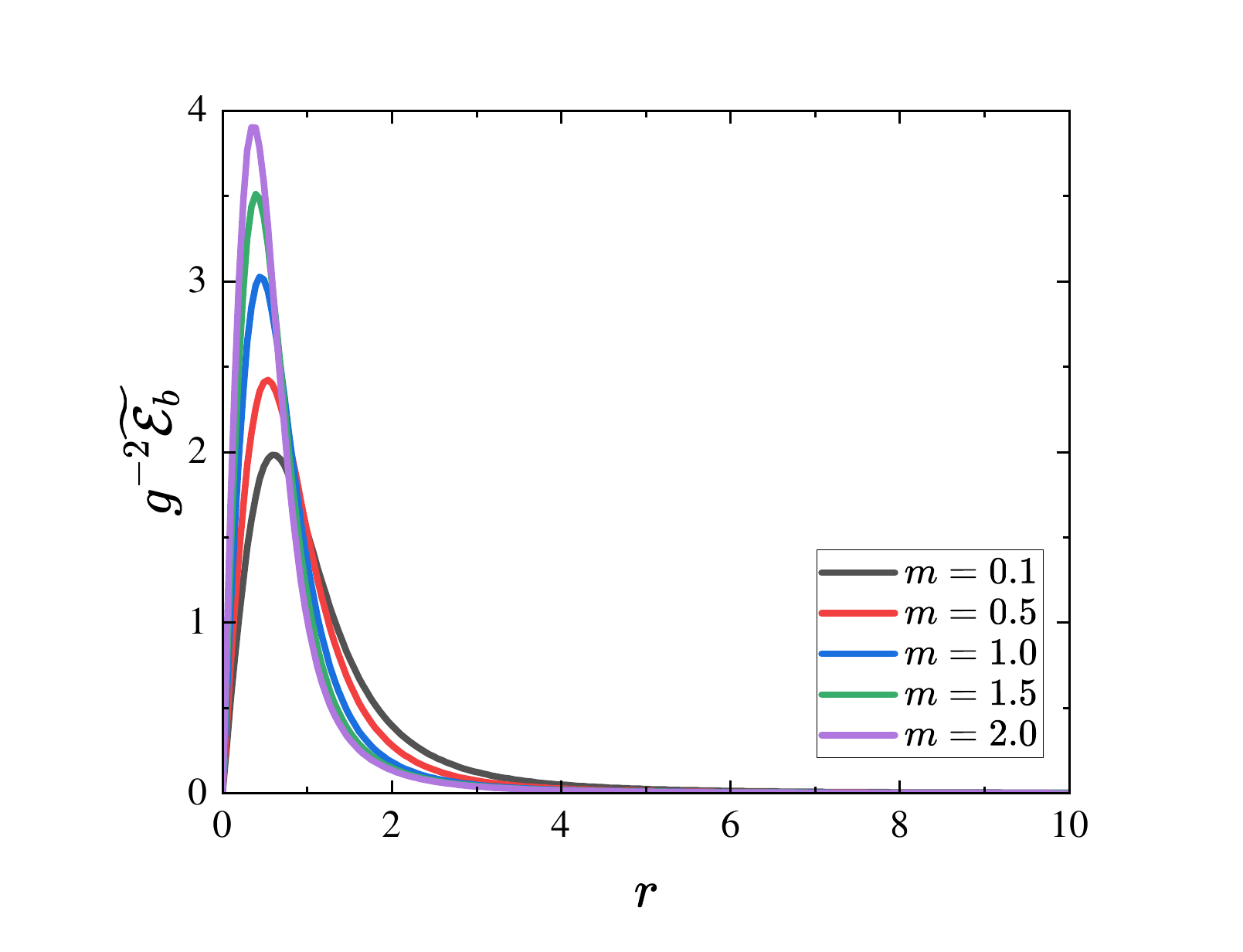}
			\subcaption{$\CP^1$ case with $Q=1\ \qty(n_1=0,n_2=1)$}
			\label{EneSol_CP1_Q1_g03}
		\end{minipage}
		\begin{minipage}[b]{0.5\linewidth}
			\centering
			\includegraphics[width=1\linewidth]{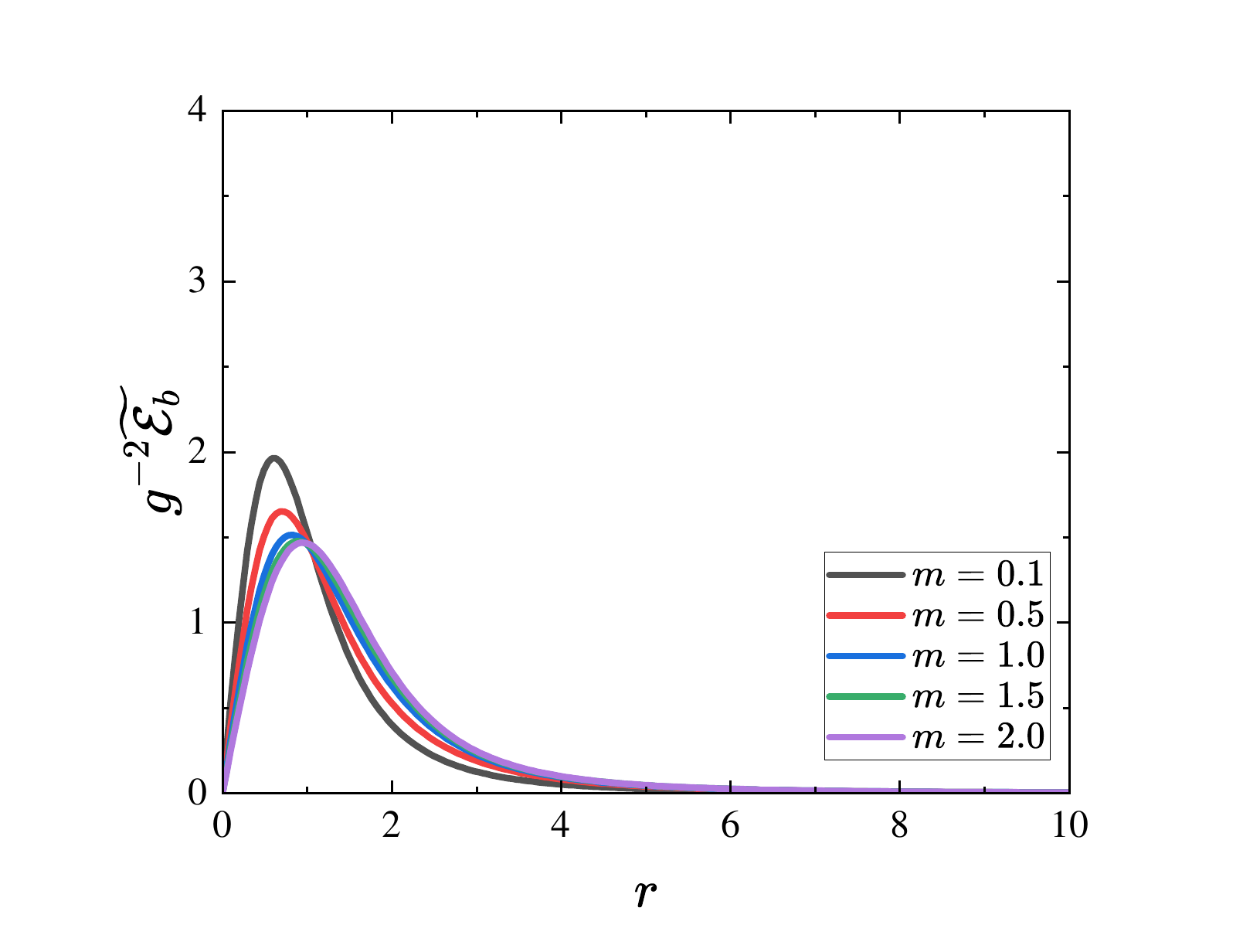}
			\subcaption{$\CP^1$ case with $Q=-1\ \qty(n_1=0,n_2=-1)$}
			\label{EneSol_CP1_Q-1_g03}
		\end{minipage}\\
		\begin{minipage}[b]{0.5\linewidth}
			\centering
			\includegraphics[width=1\linewidth]{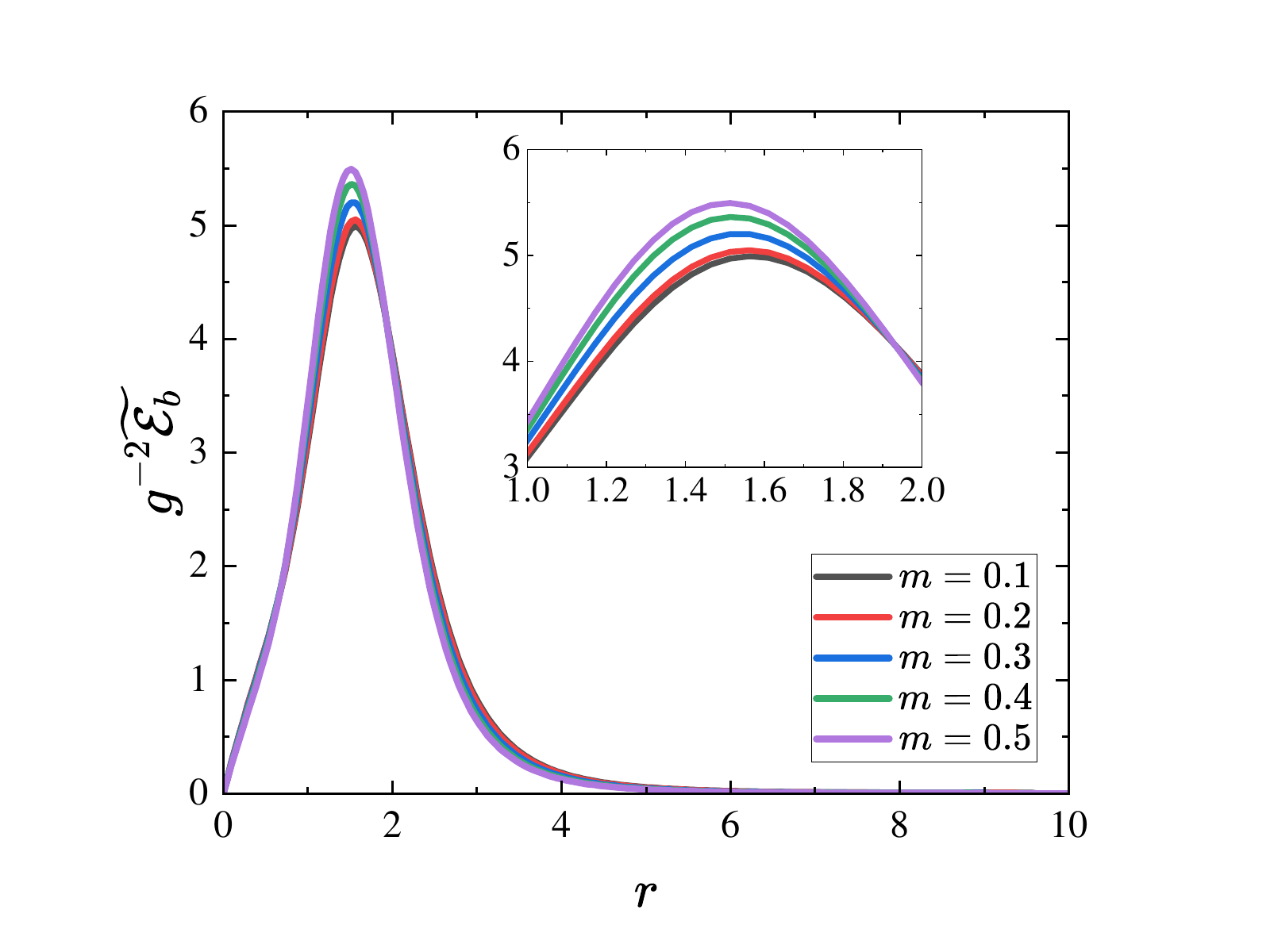}
			\subcaption{$\CP^1$ case with $Q=3\ \qty(n_1=0,n_2=3)$}
			\label{EneSol_CP1_Q3_g03}
		\end{minipage}
		\begin{minipage}[b]{0.5\linewidth}
			\centering
			\includegraphics[width=1\linewidth]{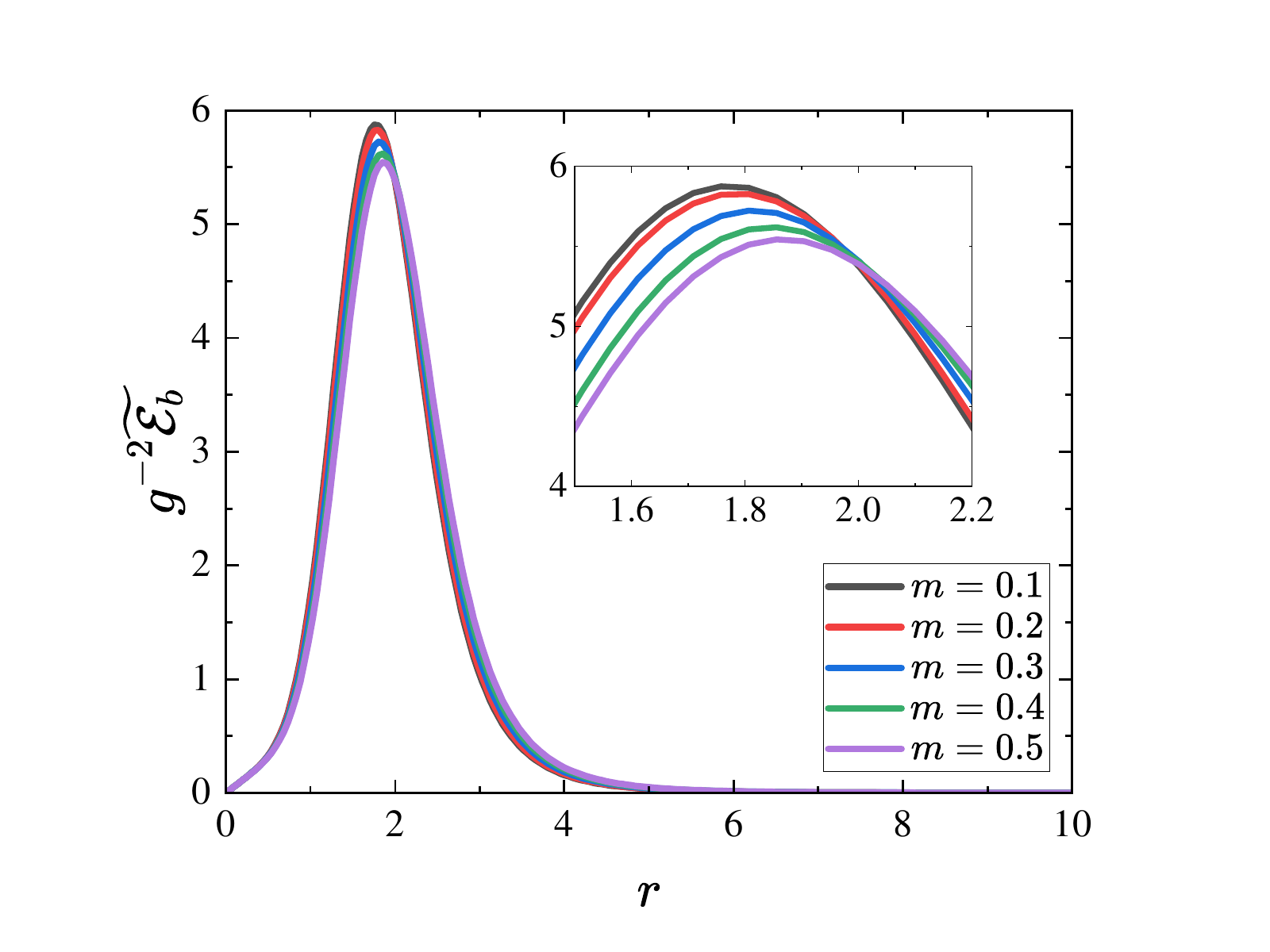}
			\subcaption{$\CP^1$ case with $Q=-3\ \qty(n_1=0,n_2=-3)$}
			\label{EneSol_CP1_Q-3_g03}
		\end{minipage}\\
		\begin{minipage}[b]{0.5\linewidth}
			\centering
			\includegraphics[width=1\linewidth]{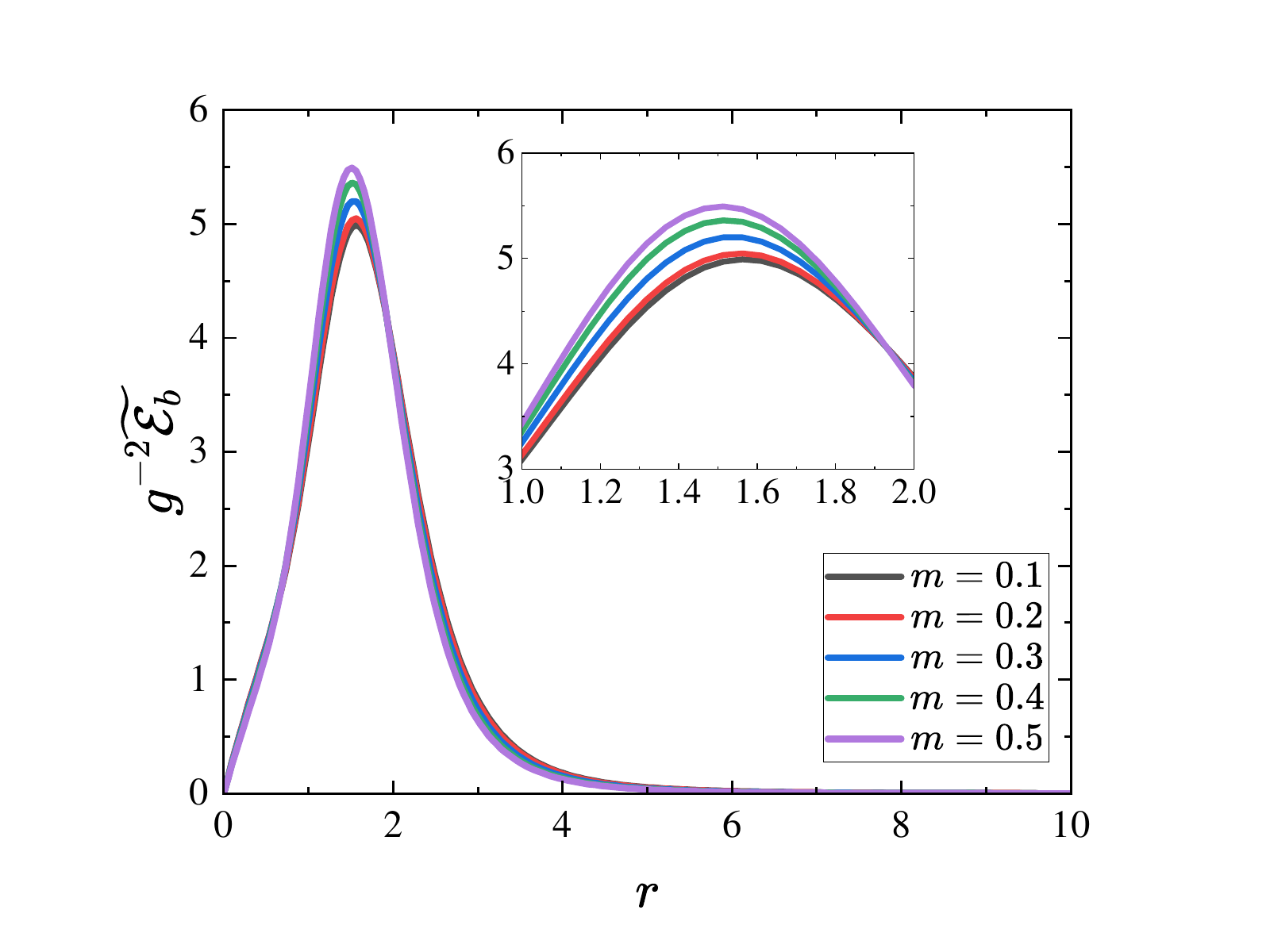}
			\subcaption{$\CP^2$ case with $Q=3\ \qty(n_1=1,n_2=3)$}
			\label{EneSol_CP2_Q3_g03}
		\end{minipage}
		\begin{minipage}[b]{0.5\linewidth}
			\centering
			\includegraphics[width=1\linewidth]{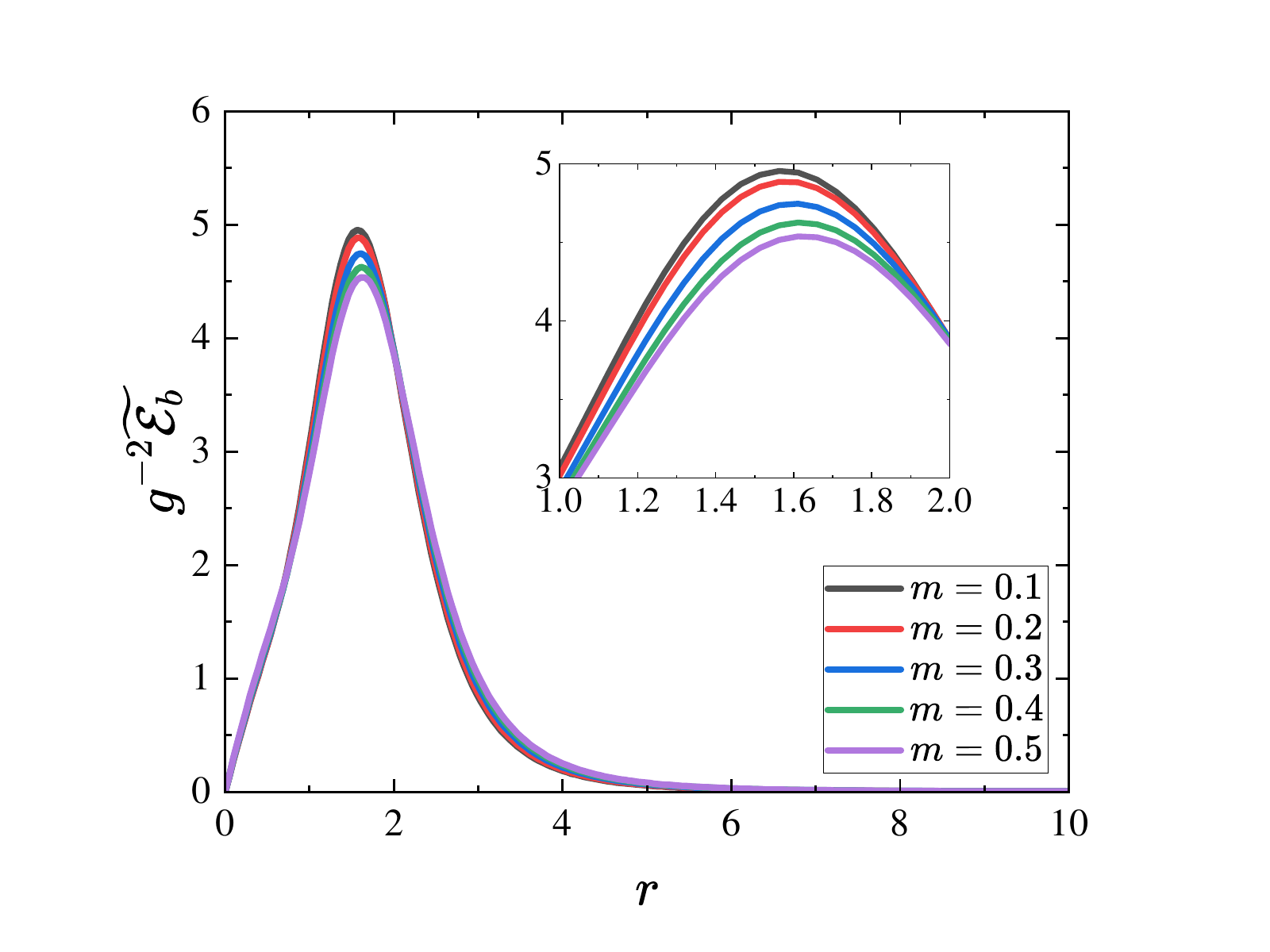}
			\subcaption{$\CP^2$ case with $Q=-3\ \qty(n_1=-1,n_2=-3)$}
			\label{EneSol_CP2_Q-3_g03}
		\end{minipage}\\
        \caption{Typical static Skyrmion energy density $g^{-2}\mathcal{E}_b$ weighted by the Jacobian $2\pi r$, $g^{-2}\widetilde{\mathcal{E}_b}\equiv 2\pi rg^{-2}\mathcal{E}_b$ for several values of the Yukawa coupling constant $m$, with $g^{-2}=0.9$. The same set of coupling values is used for both $Q>0$ and $Q<0$ to directly compare fermion-induced deformations. } 
		\label{EneSol_g03}
	\end{figure}
\end{widetext}

One can see the difference between positive topological charge cases and negative charge cases in the static energy density $g^{-2}\mathcal{E}_b$ of the Skyrmion. In Fig.~\ref{EneSol_g03}, the energy density $g^{-2}\mathcal{E}_b$ is presented for each parameter set $\qty(n_1,n_2)$. Our results in Fig.~\ref{EneSol_g03} show that increasing the Yukawa coupling constant $m$, the energy density $g^{-2}\mathcal{E}_b$ is deformed due to the backreaction. However, the deformation pattern varies depending on the sign of topological charge $Q$. For $Q>0$ cases (see left panels in Fig.~\ref{EneSol_g03}), the peak of $g^{-2}\mathcal{E}_b$ moves towards the origin, and for $Q<0$ cases (see right panels in Fig.~\ref{EneSol_g03}) it moves away from the origin. Furthermore, this difference in the deformation pattern can be interpreted as follows. For $Q>0$ cases, the backreaction acts attractively, causing the energy density $g^{-2}\mathcal{E}_b$ to concentrate at the origin. For $Q<0$ cases, the backreaction acts repulsively, displacing $g^{-2}\mathcal{E}_b$ from the origin. This is an aspect of the asymmetry. 

\begin{widetext}
	
	\begin{figure}[H]
		\begin{minipage}[b]{0.5\linewidth}
			\centering
			\includegraphics[width=1\linewidth]{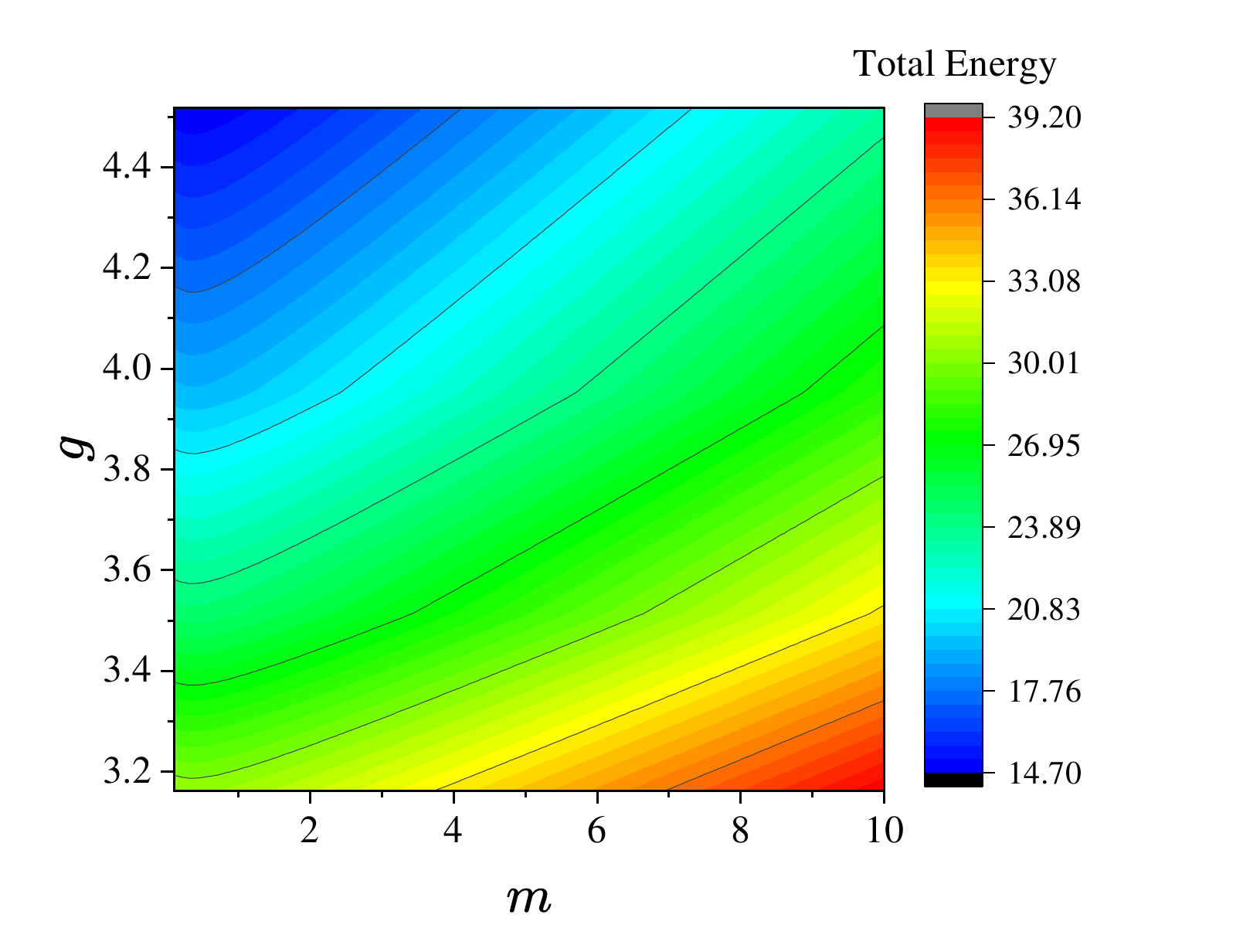}
			\subcaption{$\CP^1$ case with $Q=1\ \qty(n_1=0,n_2=1)$}
			\label{mgPhase_EneTot_CP1_Q1}
		\end{minipage}
		\begin{minipage}[b]{0.5\linewidth}
			\centering
			\includegraphics[width=1\linewidth]{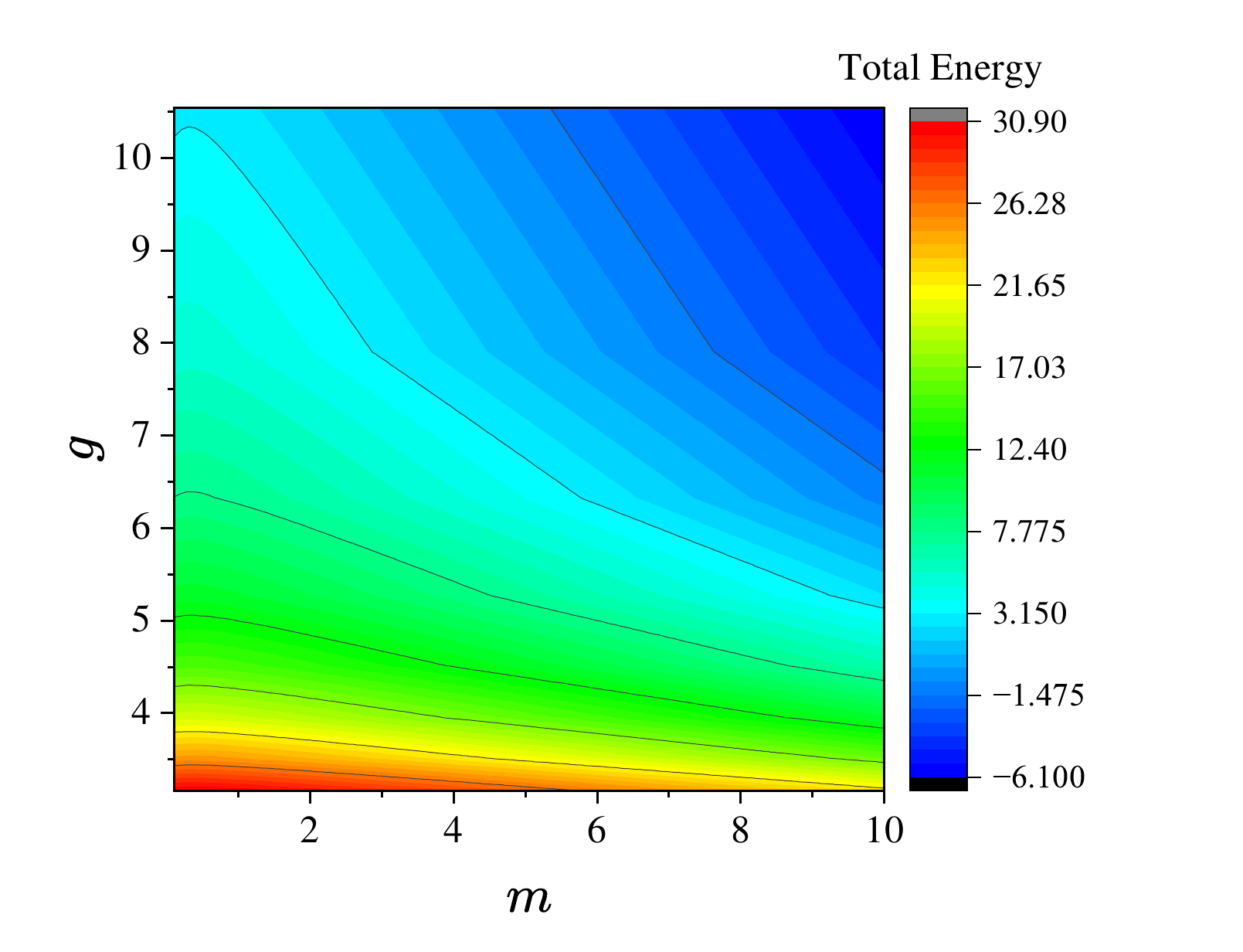}
			\subcaption{$\CP^1$ case with $Q=-1\ \qty(n_1=0,n_2=-1)$}
			\label{mgPhase_EneTot_CP1_Q-1}
		\end{minipage}\\
		\begin{minipage}[b]{0.5\linewidth}
			\centering
			\includegraphics[width=1\linewidth]{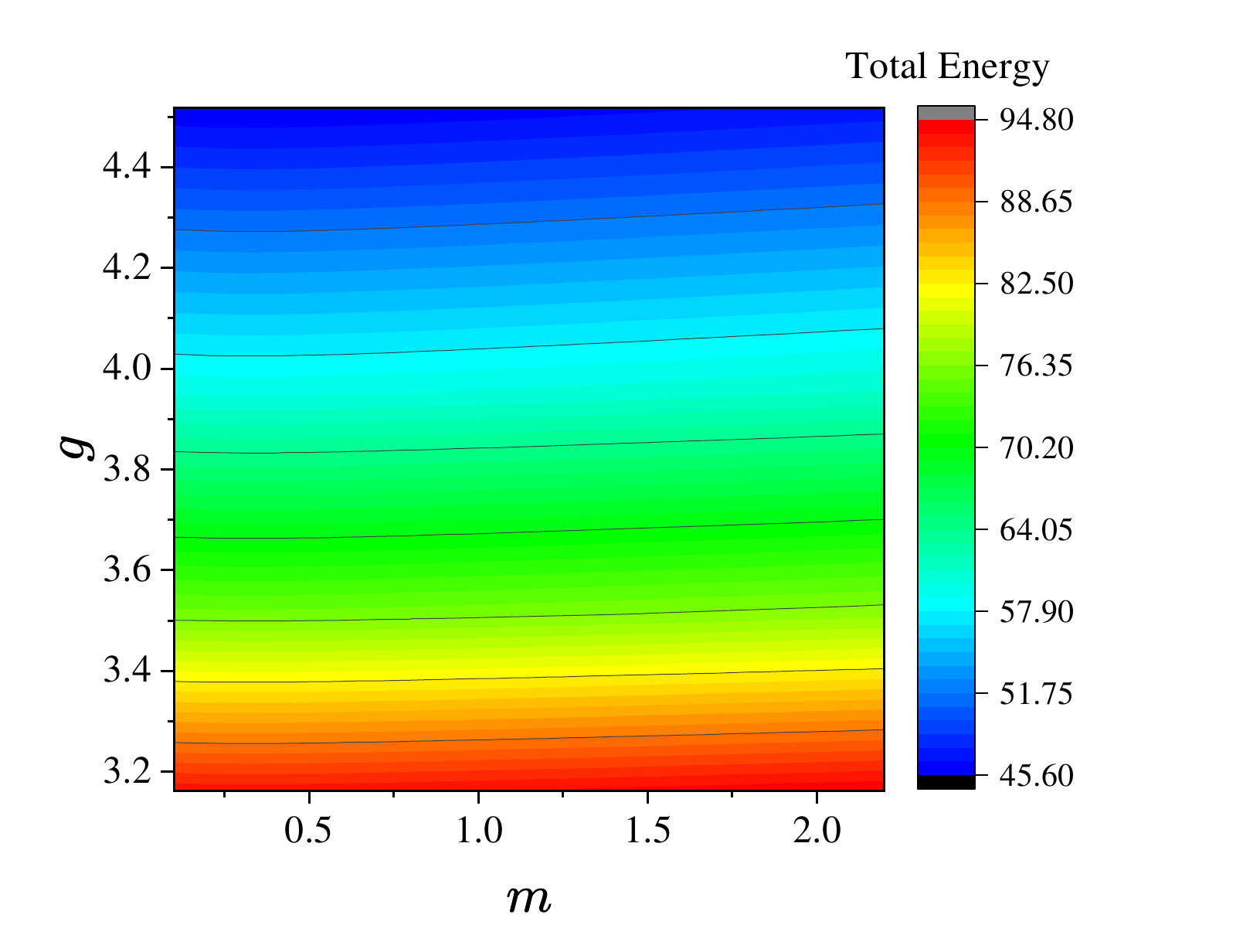}
			\subcaption{$\CP^1$ case with $Q=3\ \qty(n_1=0,n_2=3)$}
			\label{mgPhase_EneTot_CP1_Q3}
		\end{minipage}
		\begin{minipage}[b]{0.5\linewidth}
			\centering
			\includegraphics[width=1\linewidth]{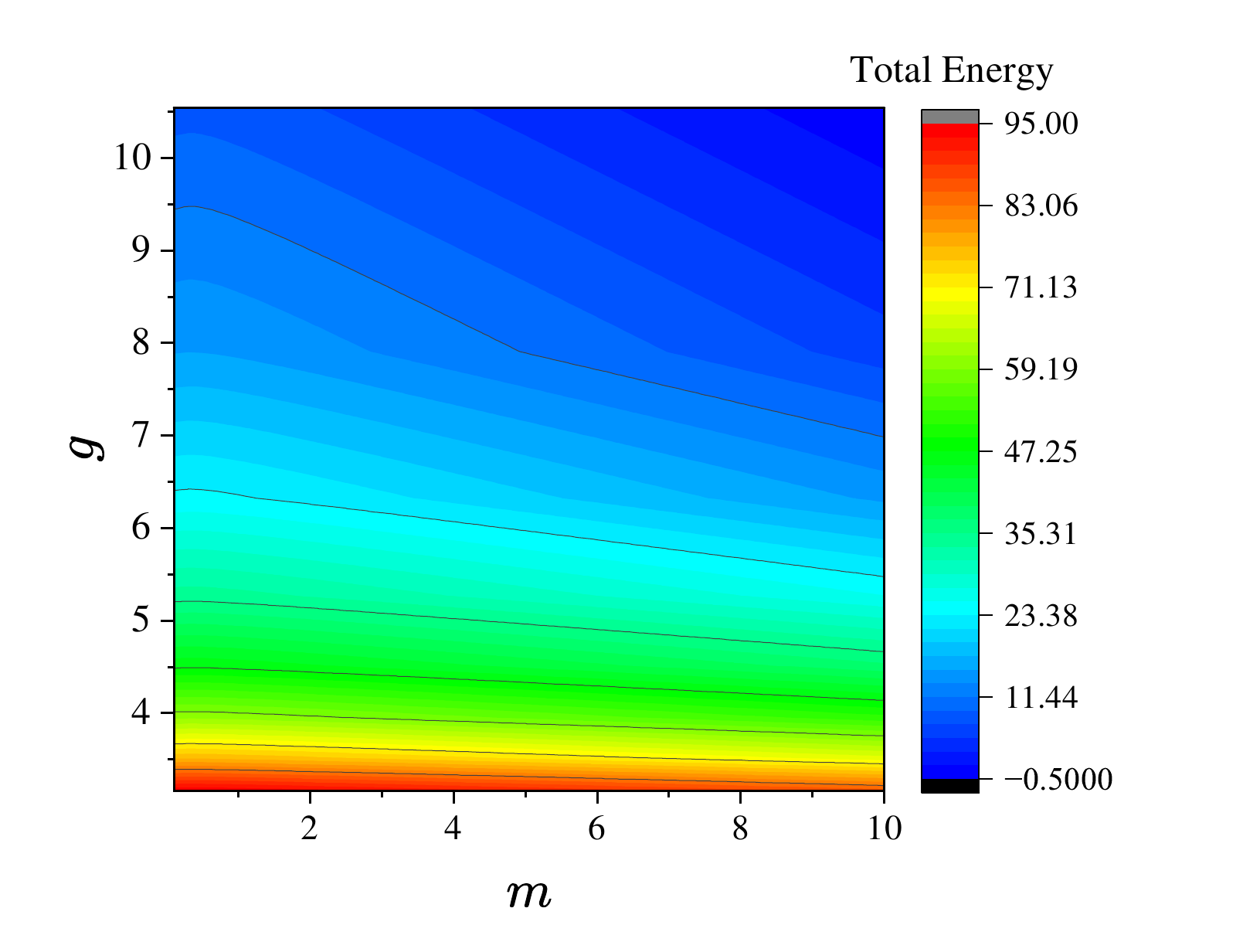}
			\subcaption{$\CP^1$ case with $Q=-3\ \qty(n_1=0,n_2=-3)$}
			\label{mgPhase_EneTot_CP1_Q-3}
		\end{minipage}\\
		\begin{minipage}[b]{0.5\linewidth}
			\centering
			\includegraphics[width=1\linewidth]{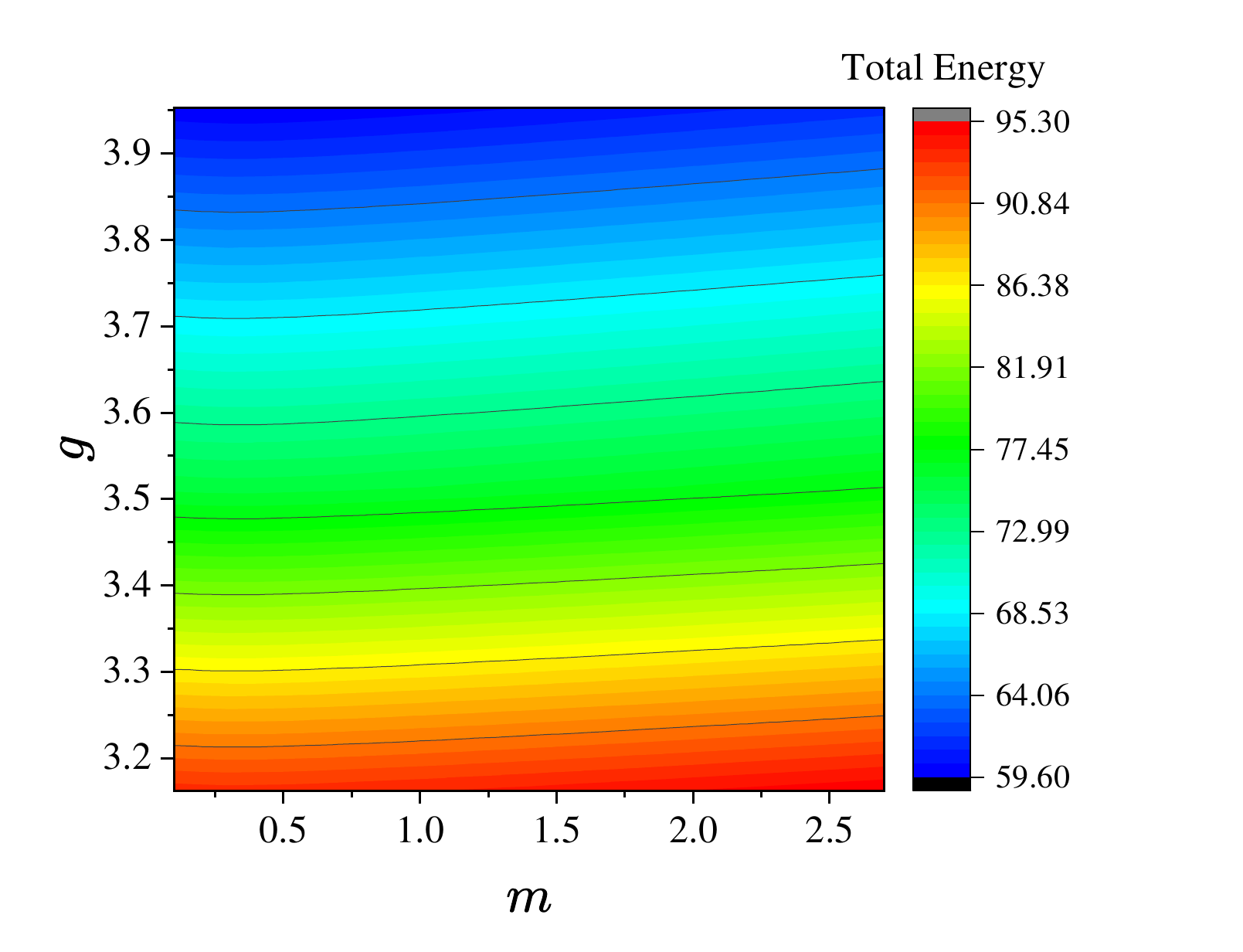}
			\subcaption{$\CP^2$ case with $Q=3\ \qty(n_1=1,n_2=3)$}
			\label{mgPhase_EneTot_CP2_Q3}
		\end{minipage}
		\begin{minipage}[b]{0.5\linewidth}
			\centering
			\includegraphics[width=1\linewidth]{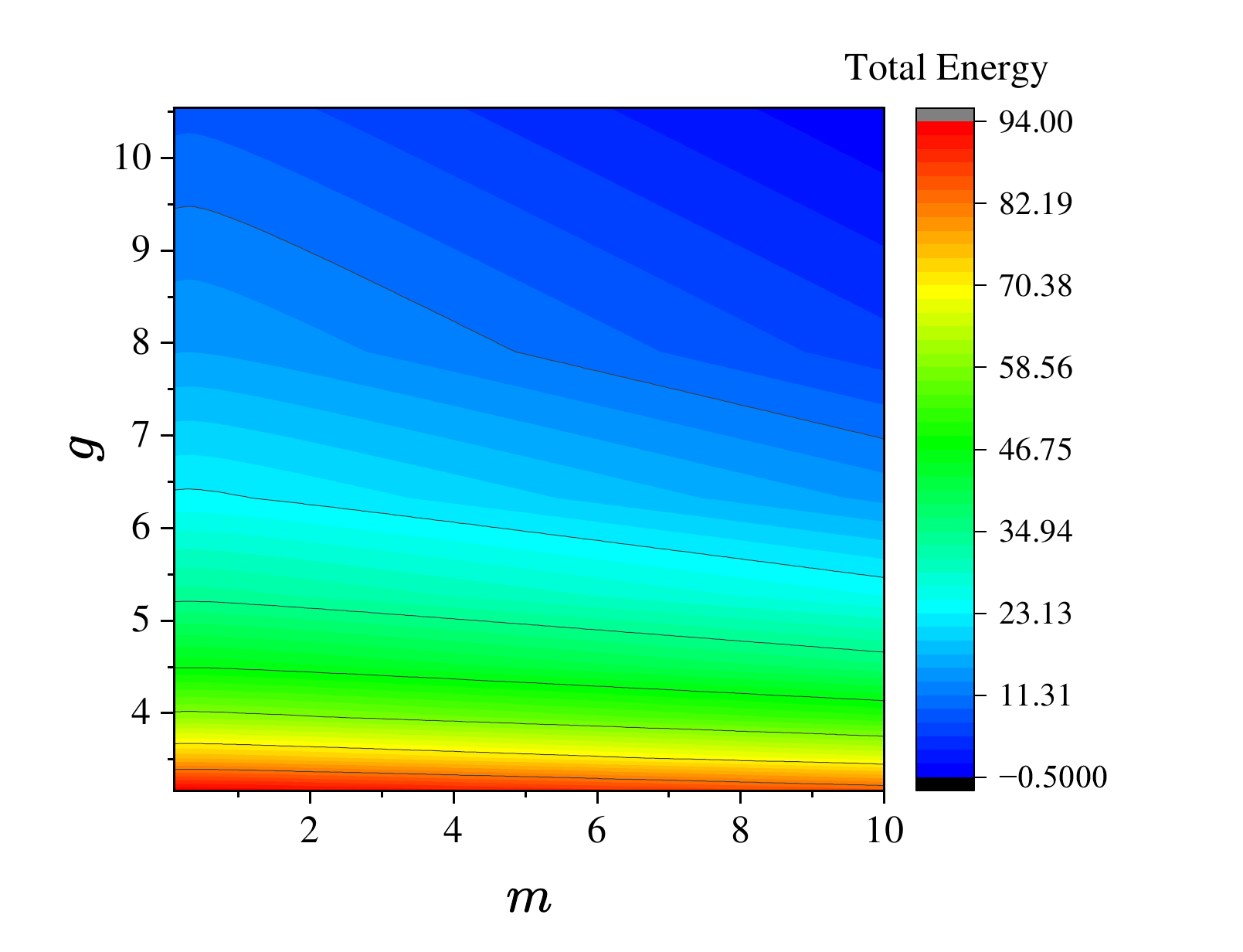}
			\subcaption{$\CP^2$ case with $Q=-3\ \qty(n_1=-1,n_2=-3)$}
			\label{mgPhase_EneTot_CP2_Q-3}
		\end{minipage}
		\caption{Total energy \eqref{eq:total_energy} in the parameter space $\qty(m,g)$. The upper four panels \ref{mgPhase_EneTot_CP1_Q1}-\ref{mgPhase_EneTot_CP1_Q-3} represent the embedded $\CP^1$ cases, and the lower two panels \ref{mgPhase_EneTot_CP2_Q3}-\ref{mgPhase_EneTot_CP2_Q-3} are genuine $\CP^2$ cases. } 
		\label{mgPhase_EneTot}
	\end{figure}
\end{widetext}

One can also find the asymmetry in the total energy of the coupled system \eqref{eq:total_energy} in the parameter space $\qty(m,g)$. The parameter dependence of the total energy is presented in Fig.~\ref{mgPhase_EneTot}. Our results exhibit a clear asymmetry of the system. For $Q>0$ cases (see left panels in Fig.~\ref{mgPhase_EneTot}), lower energy configurations are located in regions where $m$ is small and $g$ is large. This indicates that the Skyrmions with $Q>0$ possess lower energy when the backreaction from fermions is weaker. This behavior can simply be explained by the fact that increasing $g$ lowers the static energy via $g^{-2}\int\dd[2]{x}\mathcal{E}_b$. On the other hand, for $Q<0$ cases (see right panels in Fig.~\ref{mgPhase_EneTot}), lower energy configurations are located in regions where both $m$ and $g$ are large. This means that the the energy of Skyrmions with $Q<0$ decrease when the backreaction is stronger. Moreover, the Skyrmion energy is comparable to the energy-scale $\qty(\sim m)$ of fermions. 

\section{Mechanism of the Topological Charge Asymmetry}
\label{Mechanism}

We now discuss the origin of the asymmetry between configurations with topological charges $Q$ and $-Q$, even though the underlying Lagrangian is symmetric under the sign reversal of the topological charge. 

Formally, the classical action is defined as a functional of the fields. The topological charge $Q$ is not an independent variable at this level, but rather a global label that classifies distinct sectors. In the absence of topological terms that explicitly couples to $Q$, the action itself does not distinguish between configurations with opposite topological charges. Note that one can check that the reduced Lagrangian with our ansatz \eqref{Bosonicansatz} and \eqref{Fermionicansatz} is also symmetric under the transformation $n_1\to-n_1$ and $n_2\to-n_2$, namely $Q\to-Q$. 

This symmetry is also preserved at the level of the background approximation. For a fixed bosonic background $Z$ and the complex-conjugation $Z^*$, the Dirac Hamiltonian $\mathcal{H}\qty[Z]$ satisfies 
\begin{align}
	\mathcal{S}\mathcal{H}\qty[Z]\mathcal{S}^{-1}=-\mathcal{H}\qty[Z^*],\label{Sconj}
\end{align}
where the operator $\mathcal{S}$ is defined by $\mathcal{S}\equiv \sigma_1\otimes I_{3}K$ with the complex-conjugation operator $K$. This relation implies that the fermionic spectrum in the background $Z$ is mapped to that in the background $Z^*$ with all eigenvalues reversed in sign with the counterpart Dirac equation 
\begin{align}
	\mathcal{H}\qty[Z^*]\qty(\mathcal{S}\psi)&=-\varepsilon \qty(\mathcal{S}\psi).
\end{align}
Within our rotationally symmetric ansatz, the conjugation $Z\to Z^*$ corresponds to $n_1\to-n_1$ and $n_2\to-n_2$, and consequently $Q\to-Q$. Then, under the sign reversal of the topological charge $Q\to-Q$, corresponding to $Z\to Z^*$, $\psi\to \mathcal{S}\psi$ and $\varepsilon\to-\varepsilon$, the reduced Lagrangian 
\begin{align}
	\mathcal{L}_f=\psi^\dagger\qty(\varepsilon-\mathcal{H}\qty[Z])\psi
\end{align}
is transformed as 
\begin{align}
	\mathcal{L}'_f
	&=\qty(\mathcal{S}\psi)^\dagger\qty(-\varepsilon-\mathcal{H}\qty[Z^*])\qty(\mathcal{S}\psi)\nonumber\\
	&=-\mathcal{L}_f.
\end{align}
This change does not affect the classical theory. 

The situation changes once the fermionic degrees of freedom are constrained by the Dirac eigenvalue problem and incorporated self-consistently into the bosonic equations. Under the sign reversal of the topological charge $Q$ corresponding to $Z\to Z^*$ and $\psi\to \mathcal{S}\psi$, the total energy \eqref{eq:total_energy} is transformed as 
\begin{align}
	E_{\mathrm{tot}}\to E'_{\mathrm{tot}}= \int\dd[2]{x}\qty(-\psi^\dagger\mathcal{H}\psi+g^{-2}\mathcal{E}_b),
\end{align}
where the static energy density $\mathcal{E}_b$ is symmetric. As a consequence, the reduced total energy functional is no longer invariant under the sign reversal of the topological charge. Therefore, although the Lagrangian is symmetric under $Q\to-Q$, the solutions obtained after imposing the fermionic eigenvalue problem becomes inequivalent for $Q$ and $-Q$. This treatment leads to a loss of the $Q\to -Q$ symmetry, which provides the mechanism underlying the topological charge asymmetry observed in our numerical solutions.

Note that, in general, the complex-conjugate $Z\to Z^*$ does not necessarily imply $Q\to-Q$. Therefore, the above discussion relies on the specific ansatz adopted in this work \eqref{Bosonicansatz}. Nevertheless, the extension of this argument to $\CP^N$ with $N>2$ is straightforward. As long as the ansatz is chosen such that each component of $Z$ factorizes into a radial profile and an angular phase carrying the winding number, complex-conjugation reverses the corresponding windings and thus implements $Q\to-Q$ within the restricted class of configurations. Under these conditions, the mechanism discussed above generically allows a topological charge asymmetry to emerge in the numerical solutions.



\section{Summary and Outlook}
\label{SummaryandOutlook}
In this work, we investigate the $\CP^N$ Skyrmion-fermion coupled system under the assumption of rotational symmetry, focusing on how the system responds to changes in the sign of the topological charge $Q$ through the backreaction. We find that the system exhibits a distinct asymmetry with respect to the sign of $Q$. In particular, an asymmetry is found in the Skyrmion's energy density $g^{-2}\mathcal{E}_b$, which is shifted toward the origin for $Q>0$ and outward for $Q<0$. These features can be interpreted as reflecting an aspect of the backreaction: it acts attractively for $Q>0$ and repulsively for $Q<0$. Moreover, the total energy also exhibits a nontrivial asymmetry: for $Q>0$, the configuration with weaker fermion-Skyrmion interaction is energetically favored, whereas for $Q<0$, the more strongly coupled configuration is favored. This suggests that, in the backreacting fermion-soliton system, the stability condition itself may depend asymmetrically on the sign of $Q$.

We clarify the mechanism underlying the topological charge asymmetry observed in our $\CP^N$ Skyrmion-fermion system. At the level of the Lagrangian, including both bosonic and fermionic sectors, the theory is symmetric under $Q\to-Q$. This symmetry persists in the background approximation, where the fermion spectrum is reversed under complex-conjugation of the Skyrmion field. However, once the fermionic degrees of freedom are determined through the Dirac eigenvalue problem and incorporated self-consistently into the bosonic equations, the fermionic contribution changes sign while the bosonic sector remains invariant. Consequently, the self-consistent equations for $Q$ and $-Q$ become inequivalent, allowing a topological charge asymmetry to emerge.

As a future direction, it is important to analyze the dependence of the system on the potential parameter $\mu$, and to explore the existence and properties of solutions in the limit without a potential term. In addition, the asymmetry in the total energy with respect to the sign of $Q$ suggests a possible difference in the stability of the solutions. A linear stability analysis around these configurations would provide further insight into the topological charge asymmetry. 


\section*{Acknowledgments}
The authors are grateful Luiz Agostinho Ferreira, Pawel Klimas and Yakov Shnir for their many useful comments. 
This work is supported in part by JSPS KAKENHI [Grants No. JP23KJ1881 (YA), JP23K02794 (NS)] and JST SPRING [Grant Number JPMJSP2151 (SY)]. 



\appendix

\section{The coefficients in the field equations}
\label{appendixA}
The coefficients $c_i~\qty(i=1,2,\cdots,16)$ which appear in the equations \eqref{eqF} and \eqref{eqG} (and also the coupled equations \eqref{coupledODE1} and \eqref{coupledODE2}) are trigonometric polynomials which are defined as follows: 

\begin{widetext}

\begin{align}
	c_1
	&=
	4\qty{
		\cos4F 
		\qty[
			\qty(
				n_1-n_2
			) \cos2G+n_1+n_2
		]^2
		+
		\qty(
			n_1-n_2
		)^2 \cos2F \sin^22G
	}
	\nonumber\\
	&+
	8\qty[
		\qty(
			n_2^2-n_1^2
		)\cos2G
		-
		\qty(
			n_1^2+n_2^2
		)
	]
	-2r^2,\\
	c_2
	&=24
	\qty(
		n_1-n_2
	) \sin^3F\cos F\sin2G 
	\qty[
		\qty(
			n_1-n_2
		) \cos2G+n_1+n_2
	],\\
	c_3
	&=2
	\qty{
		\qty(
			n_1-n_2
		)^2 \sin2F 
		\qty(
			\cos4 G-1
		)
		-4 \sin4F 
		\qty[
			\qty(
				n_1-n_2
			) \cos2G+n_1+n_2
		]^2
	},\\
	c_4
	&=
	2\sin F \cos F \cos2 F 
		\qty{
			3 
			\qty(
				n_1-n_2
			) \sin^2F 
			\qty[
				4 
				\qty(
					n_1+n_2
				) \cos2G+
				\qty(
					n_1-n_2
				) \cos4G
			]
			-5n_1^2+18 n_1 n_2-5 n_2^2
		}\nonumber\\
	&+2\sin F \cos F  
		\qty(
			n_1-n_2
		) \sin^2F 
		\qty[
			28 
			\qty(
				n_1+n_2
			) \cos2G+9 
			\qty(
				n_1-n_2
			) \cos4G
		]\nonumber\\
	&-\frac{3}{2} \sin F \cos F  
		\qty(
			3n_1^2+2 n_1 n_2+3 n_2^2
		) \cos4F
	+\frac{1}{2} \sin F \cos F 
		\qty(
			29 n_1^2-66 n_1 n_2+29 n_2^2
		)
	+2r^2 \sin F \cos F. \\
	c_5
	&=16
	\qty(
		n_1-n_2
	) 
	\qty[
		2 
		\qty(
			n_1+n_2
		) \sin^22F \sin2G+
		\qty(
			n_1-n_2
		) \sin^2F 
		\qty(
			2 \cos2 F+1
		) \sin4 G
	],\\
	c_6
	&= 
	-4\qty{
		\cos4F 
		\qty[
			\qty(
				n_1-n_2
			) \cos2 G+n_1+n_2
		]^2
		+
		\qty(
			n_1-n_2
		)^2 \cos2F \sin^22 G
	}
	\nonumber\\
	&-8\qty[
		\qty(
			n_2^2-n_1^2
		)\cos2G
		-
		\qty(
			n_1^2+n_2^2
		)
	]
	-2r^2,\\
	c_7
	&=-24
	\qty(
		n_1-n_2
	) \sin^3F \cos F \sin2 G 
	\qty[
		(
			n_1-n_2
		) \cos2G+n_1+n_2
	],\\
	c_8
	&=\frac{1}{16}\sin4F 
	\qty[
		4 
		\qty(
			n_1-n_2
		) 
		\qty(
			n_1+n_2
		) \cos2G+3 n_1^2+2 n_1 n_2+3 n_2^2
	]\nonumber\\
	&-\frac{1}{8} 
	\qty(
		n_1-n_2
	)^2 
	\qty(
		4 \sin^3F \cos F \cos4G-\sin2F
	)\nonumber\\
	&+\frac{\mu^2}{64} r^2 
	\qty[
		8\sin^3F\cos F \cos4G+2 \sin2F 
		\qty(
			4 \cos2G-5
		)+\sin4F 
		\qty(
			4 \cos2G-3
		)
	],\\
	c_9
	&=
	-2 \sin^2F \sin^2F 
		\qty[
			8
			\qty(
				n_1^2-n_2^2
			) \cos^2F \cos2G+
			\qty(
				n_1-n_2
			)^2 
			\qty(
				\cos2F-7
			) \cos4G
		]\nonumber\\
	&-2 \sin^2F \sin^2F 
		\qty[
			\qty(
				3n_1^2+2 n_1 n_2+3 n_2^2
			) \cos2F+11 n_1^2-14 n_1 n_2+11 n_2^2
		]
	-2r^2 \sin^2F,\\
	c_{10}
	&=24
	\qty(
		n_1-n_2
	) \sin^3F\cos F\sin2G 
	\qty[
		\qty(
			n_1-n_2
		) \cos2G+n_1+n_2
	],\\
	c_{11}
	&=4
	\qty(
		n_1-n_2
	) \sin^4F 
	\qty[
		4 
		\qty(
			n_1+n_2
		) \cos^2F \sin2G+
		\qty(
			n_1-n_2
		) 
		\qty(
			\cos2F-7
		) \sin4G
	],\\
	c_{12}
	&=4
	\qty(
		n_1-n_2
	) \sin^2F 
	\qty[
		2 
		\qty(
			n_1+n_2
		) 
		\qty(
			2 \cos2F-1
		) \sin2G+
		\qty(
			n_1-n_2
		) 
		\qty(
			2 \cos2F+1
		) \sin4G
	],\\
	c_{13}
	&=
	-24\sin^3F \cos F  \cos2F 
		\qty[
			\qty(
				n_1-n_2
			) \cos2G+n_1+n_2
		]^2\nonumber\\
	&-16\sin^3F \cos F 
		\qty(
			n_1-n_2
		) 
		\qty(
			n_1+n_2
		) \cos2G
		+60\sin^3F \cos F 
		\qty(
			n_1-n_2
		)^2 \cos4G\nonumber\\
	&+4\sin^3F \cos F 
		\qty(
			-19 n_1^2+30 n_1 n_2-19 n_2^2
		)-2 r^2 \sin2F,\\
	c_{14}
	&=2
	\qty(
		n_1-n_2
	) \sin^4F 
	\qty[
		8 
		\qty(
			n_1+n_2
		) \cos^2F \cos2G+
		\qty(
			n_1-n_2
		) 
		\qty(
			\cos2F-7
		) \cos4G
	]\nonumber\\
	&-4 \sin^2F
	\qty[
		\qty(
			3 n_1^2+2 n_1 n_2+3 n_2^2
		) \cos4F+16 
		\qty(
			n_1-n_2
		)^2 \cos2F-19 n_1^2+30 n_1 n_2-19 n_2^2+4r^2
	],\\
	c_{15}
	&=-24
	\qty(
		n_1-n_2
	) \sin^3F\cos F \sin2 G 
	\qty[
		\qty(
			n_1-n_2
		) \cos2G+n_1+n_2
	],\\
	c_{16}
	&=-\frac{1}{2}
	\qty(
		n_1-n_2
	) \sin^2F 
	\qty[
		\qty(
			n_2-n_1
		) \sin^2F \sin4G+2 
		\qty(
			n_1+n_2
		) \cos^2F \sin2G
	]\nonumber\\
	&-\frac{\mu^2}{8} r^2 \sin^2F 
	\qty[
		\sin^2F \sin4G+
		\qty(
			\cos2F+3
		) \sin2G
	].
\end{align}
\end{widetext}
Note that these coefficients themselves are invariant under $n_1\to-n_1$ and $n_2\to-n_2$, namely $Q\to-Q$. 

\section{The Plane Wave Basis}
\label{appendixB}
In this section, we provide the plane wave basis $\Phi_\kappa^{(p)}$ which we use to solve the secular equation \eqref{Secular} for the $\CP^2$ case~\cite{Amari:2019tgs}. The basis is the eigenvector of the Hamiltonian of the massive free fermions $\mathcal{H}^0=\hat{\gamma}^0\qty(-i\hat{\gamma}^j\partial_j+m\hat{\Sigma}_\infty)$. The explicit form can be summarized as follows: 

\begin{widetext}

\begin{align}
    \Phi_\kappa^{(u)}&=\mathcal{N}_i^{(u)}
	\mqty
	(
		\zeta_{+}^{(u)}J_{\kappa-\frac{1}{2}-\frac{n_1+n_2}{3}}(k_i^{(u)}r)\exp\qty[i\qty(\kappa-\frac{1}{2}-\frac{n_1+n_2}{3})\varphi]\\
		\zeta_{-}^{(u)}J_{\kappa+\frac{1}{2}-\frac{n_1+n_2}{3}}(k_i^{(u)}r)\exp\qty[i\qty(\kappa+\frac{1}{2}-\frac{n_1+n_2}{3})\varphi]
	)
	\otimes
	\mqty
	(
		1\\
		0\\
		0
	),\\
    \Phi_\kappa^{(d)}&=\mathcal{N}_i^{(d)}
	\mqty
	(
		\zeta_{+}^{(d)}J_{\kappa-\frac{1}{2}-\frac{-2n_1+n_2}{3}}(k_i^{(d)}r)\exp\qty[i\qty(\kappa-\frac{1}{2}-\frac{-2n_1+n_2}{3})\varphi]\\
		\zeta_{-}^{(d)}J_{\kappa+\frac{1}{2}-\frac{-2n_1+n_2}{3}}(k_i^{(d)}r)\exp\qty[i\qty(\kappa+\frac{1}{2}-\frac{-2n_1+n_2}{3})\varphi]
	)
	\otimes
	\mqty
	(
		0\\
		1\\
		0
	),\\
    \Phi_\kappa^{(s)}&=\mathcal{N}_i^{(s)}
	\mqty
	(
		\zeta_{-}^{(s)}J_{\kappa-\frac{1}{2}-\frac{n_1-2n_2}{3}}(k_i^{(s)}r)\exp\qty[i\qty(\kappa-\frac{1}{2}-\frac{n_1-2n_2}{3})\varphi]\\
		\zeta_{+}^{(s)}J_{\kappa+\frac{1}{2}-\frac{n_1-2n_2}{3}}(k_i^{(s)}r)\exp\qty[i\qty(\kappa+\frac{1}{2}-\frac{n_1-2n_2}{3})\varphi]
	)
	\otimes
	\mqty
	(
		0\\
		0\\
		1
	),
\end{align}
\end{widetext}
where $J_{l^{\qty(p)}}\qty(k_i^{(p)}r)$ are standard Bessel functions with the integer order $l^{\qty(p)}$. The coefficients $\zeta$ are given by 
\begin{align}
	\zeta^{(p)}_{+}&=
    \begin{cases}
    \ \ \ \ \ \ \ 1&\qty(\omega_i^{\qty(p)}>0)\\
    \frac{k_i^{\qty(p)}}{\abs{\omega_i^{\qty(p)}}+m}&\qty(\omega_i^{\qty(p)}<0)
    \end{cases}\nonumber\\
    \zeta^{(p)}_{-}&=
    \begin{cases}
    -\frac{k_i^{\qty(p)}}{\omega_i^{\qty(p)}+m}&\qty(\omega_i^{\qty(p)}>0)\\
    \ \ \ \ \ \ \ 1&\qty(\omega_i^{\qty(p)}<0)
    \end{cases}
\end{align}
where $\omega_i^{\qty(p)}=\pm\sqrt{\qty(k_i^{\qty(p)})^2+m^2}$ are the vacuum energy eigenvalues.

We can construct the bases in a cylinder of radius $R$. This radious is ideally infinite, however for numerical computation, it must take a finite value. Imposing boundary conditions
\begin{align}
    J_{l^{\qty(p)}}\qty(k_i^{\qty(p)}R)&=0
\end{align}
for $l^{\qty(p)}=\kappa-\frac{1}{2}-\frac{n_1+n_2}{3},\ \kappa-\frac{1}{2}-\frac{-2n_1+n_2}{3},\ \text{or}\ \kappa-\frac{1}{2}-\frac{n_1-2n_2}{3}$, we obtain some sets of discretized wave numbers $k_i^{\qty(p)},p=\qty(u,d,s)$. The orthogonality conditions are given by 
\begin{align}
    &\int_0^{R}\dd{r}r J_{l^{\qty(p)}}\qty(k_i^{\qty(p)}r)J_{l^{\qty(p)}}\qty(k_j^{\qty(p)}r)\nonumber\\
    &=\int_0^{R}\dd{r}r J_{{l^{\qty(p)}}\pm1}\qty(k_i^{\qty(p)}r)J_{{l^{\qty(p)}}\pm1}\qty(k_j^{\qty(p)}r)\nonumber\\
    &=\delta_{ij}\frac{R^2}{2}\qty[J_{{l^{\qty(p)}}\pm1}\qty(k_i^{\qty(p)}R)]^2.
\end{align}
Then, the normalization constants $\mathcal{N}_i^{(p)}$ are determined as
\begin{align}
    \mathcal{N}_i^{(p)}&=\qty[\frac{2\pi R^2\abs{\omega_i^{\qty(p)}}}{\abs{\omega_i^{\qty(p)}}+m}\qty(J_{l^{\qty(p)}}\qty(k_i^{\qty(p)}R))^2]^{-\frac{1}{2}}.
\end{align}
In this setup, the plane waves form an orthonormal basis.

\bibliography{ref}

@article{Abanov:2001iz,
    author = "Abanov, A. G. and Wiegmann, P. B.",
    title = "{On the correspondence between fermionic number and statistics of solitons}",
    eprint = "hep-th/0105213",
    archivePrefix = "arXiv",
    doi = "10.1088/1126-6708/2001/10/030",
    journal = "JHEP",
    volume = "10",
    pages = "030",
    year = "2001"
}

@article{Abanov:2004ci,
    author = "Abanov, Alexander G. and Braverman, Maxim",
    title = "{Topological calculation of the phase of the determinant of a non selfadjoint elliptic operator}",
    eprint = "math-ph/0401037",
    archivePrefix = "arXiv",
    doi = "10.1007/s00220-005-1394-6",
    journal = "Commun. Math. Phys.",
    volume = "259",
    pages = "287--305",
    year = "2005"
}

@article{Akagi:2021lva,
    author = "Akagi, Yutaka and Amari, Yuki and Gudnason, Sven Bjarke and Nitta, Muneto and Shnir, Yakov",
    title = "{Fractional Skyrmion molecules in a $\mathbb{C}P^{N-1}$ model}",
    eprint = "2107.13777",
    archivePrefix = "arXiv",
    primaryClass = "hep-th",
    doi = "10.1007/JHEP11(2021)194",
    journal = "JHEP",
    volume = "11",
    pages = "194",
    year = "2021"
}

@article{Amado:2014waa,
    author = "Amado, Andr{\'e} and Mohammadi, Azadeh",
    title = "{Coupled fermion{\textendash}kink system in Jackiw{\textendash}Rebbi model}",
    eprint = "1406.1459",
    archivePrefix = "arXiv",
    primaryClass = "hep-th",
    doi = "10.1140/epjc/s10052-017-5044-x",
    journal = "Eur. Phys. J. C",
    volume = "77",
    number = "7",
    pages = "465",
    year = "2017"
}

@article{Amari:2016ynl,
    author = "Amari, Yuki and Klimas, Pawel and Sawado, Nobuyuki",
    title = "{Collective coordinate quantization, spin statistics of the solitons in the $\mathbb{C}P^N$ Skyrme-Faddeev model}",
    eprint = "1604.06125",
    archivePrefix = "arXiv",
    primaryClass = "hep-th",
    doi = "10.1103/PhysRevD.94.025032",
    journal = "Phys. Rev. D",
    volume = "94",
    number = "2",
    pages = "025032",
    year = "2016"
}

@article{Amari:2019tgs,
    author = "Amari, Yuki and Iida, Masaya and Sawado, Nobuyuki",
    title = "{Statistical Nature of Skyrme-Faddeev Models in 2+1 Dimensions and Normalizable Fermions}",
    eprint = "1910.10431",
    archivePrefix = "arXiv",
    primaryClass = "hep-th",
    doi = "10.1134/S0040577919090010",
    journal = "Theor. Math. Phys.",
    volume = "200",
    number = "3",
    pages = "1253--1268",
    year = "2019"
}

@article{Amari:2022boe,
    author = "Amari, Yuki and Akagi, Yutaka and Gudnason, Sven Bjarke and Nitta, Muneto and Shnir, Yakov",
    title = "{CP2 skyrmion crystals in an SU(3) magnet with a generalized Dzyaloshinskii-Moriya interaction}",
    eprint = "2204.01476",
    archivePrefix = "arXiv",
    primaryClass = "cond-mat.str-el",
    doi = "10.1103/PhysRevB.106.L100406",
    journal = "Phys. Rev. B",
    volume = "106",
    number = "10",
    pages = "L100406",
    year = "2022"
}

@article{Amari:2023gjq,
    author = "Amari, Yuki and Sawado, Nobuyuki and Yamamoto, Shintaro",
    title = "{Instanton size dependence on fermion energy spectra in a \ensuremath{\mathbb{C}}P$^{2}$ fermionic sigma model}",
    doi = "10.1088/1742-6596/2667/1/012024",
    journal = "J. Phys. Conf. Ser.",
    volume = "2667",
    number = "1",
    pages = "012024",
    year = "2023"
}

@article{Amari:2024adu,
    author = "Amari, Yuki and Eto, Minoru and Nitta, Muneto",
    title = "{Topological solitons stabilized by a background gauge field and soliton-anti-soliton asymmetry}",
    eprint = "2403.06778",
    archivePrefix = "arXiv",
    primaryClass = "hep-th",
    reportNumber = "YGHP-24-02",
    doi = "10.1007/JHEP11(2024)127",
    journal = "JHEP",
    volume = "11",
    pages = "127",
    year = "2024"
}

@article{Amari:2024rpm,
    author = "Amari, Yuki and Sawado, Nobuyuki and Yamamoto, Shintaro",
    title = "{Spectral flow of fermions in the {\ensuremath{\mathbb{C}}}P$^{2}$ (anti-)instanton, and the sphaleron with vanishing topological charge}",
    eprint = "2403.10149",
    archivePrefix = "arXiv",
    primaryClass = "hep-th",
    doi = "10.1007/JHEP06(2024)057",
    journal = "JHEP",
    volume = "06",
    pages = "057",
    year = "2024"
}

@article{Amari:2024pnw,
    author = "Amari, Yuki and Antsipovich, Sergei and Nitta, Muneto and Shnir, Yakov",
    title = "{Isospinning CP2 solitons}",
    doi = "10.1103/PhysRevD.110.085008",
    journal = "Phys. Rev. D",
    volume = "110",
    number = "8",
    pages = "085008",
    year = "2024"
}

@misc{Amari:2025rgt,
    author = "Amari, Yuki and Sawado, Nobuyuki and Yamamoto, Shintaro",
    title = "{$\mathbb{C}\mathrm{P}^2$ Skyrmion with Fermion Backreaction}",
    eprint = "2512.07337",
    archivePrefix = "arXiv",
    primaryClass = "hep-th",
    month = "12",
    year = "2025"
}

@article{Atiyah:1963zz,
    author = "Atiyah, M. F. and Singer, I. M.",
    title = "{The index of elliptic operators on compact manifolds}",
    doi = "10.1090/S0002-9904-1963-10957-X",
    journal = "Bull. Am. Math. Soc.",
    volume = "69",
    pages = "422--433",
    year = "1969"
}

@article{atiyah_patodi_singer_1975, 
    title={Spectral asymmetry and Riemannian Geometry. I}, 
    volume={77}, 
    DOI={10.1017/S0305004100049410}, 
    number={1}, 
    journal={Mathematical Proceedings of the Cambridge Philosophical Society}, 
    publisher={Cambridge University Press}, 
    author={Atiyah, M. F. and Patodi, V. K. and Singer, I. M.}, 
    year={1975}, 
    pages={43–69}
}

@article{Blas:2022oxs,
    author = "Blas, H. and Monsalve, J. J. and Quica{\~n}o, R. and Pereira, J. R. V.",
    title = "{Majorana zero mode-soliton duality and in-gap and BIC bound states in modified Toda model coupled to fermion}",
    eprint = "2207.01161",
    archivePrefix = "arXiv",
    primaryClass = "hep-th",
    doi = "10.1007/JHEP09(2022)082",
    journal = "JHEP",
    volume = "09",
    pages = "082",
    year = "2022"
}

@article{Burnier:2006za,
    author = "Burnier, Yannis",
    title = "{Anomalous fermion number nonconservation: Paradoxes in the level crossing picture}",
    eprint = "hep-ph/0609028",
    archivePrefix = "arXiv",
    doi = "10.1103/PhysRevD.74.105013",
    journal = "Phys. Rev. D",
    volume = "74",
    pages = "105013",
    year = "2006"
}

@article{Callan:1982ac,
    author = "Callan, Jr., Curtis G.",
    title = "{Monopole Catalysis of Baryon Decay}",
    reportNumber = "LPTENS-82-20",
    doi = "10.1016/0550-3213(83)90677-6",
    journal = "Nucl. Phys. B",
    volume = "212",
    pages = "391--400",
    year = "1983"
}

@article{DAdda:1978vbw,
    author = "D'Adda, A. and Luscher, M. and Di Vecchia, P.",
    title = "{A 1/n Expandable Series of Nonlinear Sigma Models with Instantons}",
    reportNumber = "Print-78-0885 (BOHR INST.)",
    doi = "10.1016/0550-3213(78)90432-7",
    journal = "Nucl. Phys. B",
    volume = "146",
    pages = "63--76",
    year = "1978"
}

@article{Delsate:2011aa,
    author = "Delsate, Terence and Sawado, Nobuyuki",
    title = "{Localizing modes of massive fermions and a U(1) gauge field in the inflating baby-skyrmion branes}",
    eprint = "1112.2714",
    archivePrefix = "arXiv",
    primaryClass = "gr-qc",
    doi = "10.1103/PhysRevD.85.065025",
    journal = "Phys. Rev. D",
    volume = "85",
    pages = "065025",
    year = "2012"
}

@article{Dzhunushaliev:2024kti,
    author = "Dzhunushaliev, Vladimir and Folomeev, Vladimir and Kunz, Jutta and Shnir, Yakov",
    title = "{Gravitating Skyrmions with localized fermions}",
    eprint = "2407.17504",
    archivePrefix = "arXiv",
    primaryClass = "hep-th",
    doi = "10.1140/epjc/s10052-025-14074-4",
    journal = "Eur. Phys. J. C",
    volume = "85",
    number = "4",
    pages = "391",
    year = "2025"
}

@article{Ferreira:2010jb,
    author = "Ferreira, L. A. and Klimas, P.",
    title = "{Exact vortex solutions in a $CP^N$ Skyrme-Faddeev type model}",
    eprint = "1007.1667",
    archivePrefix = "arXiv",
    primaryClass = "hep-th",
    doi = "10.1007/JHEP10(2010)008",
    journal = "JHEP",
    volume = "10",
    pages = "008",
    year = "2010"
}

@article{Gani:2010pv,
    author = "Gani, V. A. and Ksenzov, V. G. and Kudryavtsev, A. E.",
    title = "{Example of a self-consistent solution for a fermion on domain wall}",
    eprint = "1001.3305",
    archivePrefix = "arXiv",
    primaryClass = "hep-th",
    doi = "10.1134/S1063778810110104",
    journal = "Phys. Atom. Nucl.",
    volume = "73",
    pages = "1889--1892",
    year = "2010"
}

@article{Gani:2022ity,
    author = "Gani, Vakhid A. and Gorina, Anastasia and Perapechka, Ilya and Shnir, Yakov",
    title = "{Remarks on sine-Gordon kink{\textendash}fermion system: localized modes and scattering}",
    eprint = "2205.13437",
    archivePrefix = "arXiv",
    primaryClass = "hep-th",
    doi = "10.1140/epjc/s10052-022-10707-0",
    journal = "Eur. Phys. J. C",
    volume = "82",
    number = "8",
    pages = "757",
    year = "2022"
}

@article{GOBEL20211,
title = {Beyond skyrmions: Review and perspectives of alternative magnetic quasiparticles},
journal = {Physics Reports},
volume = {895},
pages = {1-28},
year = {2021},
note = {Beyond skyrmions: Review and perspectives of alternative magnetic quasiparticles},
issn = {0370-1573},
doi = {https://doi.org/10.1016/j.physrep.2020.10.001},
author = {Börge Göbel and Ingrid Mertig and Oleg A. Tretiakov}
}

@article{Golo:1978de,
    author = "Golo, V. L. and Perelomov, A. M.",
    title = "{Solution of the Duality Equations for the Two-Dimensional SU(N) Invariant Chiral Model}",
    reportNumber = "ITEP-62-1978",
    doi = "10.1016/0370-2693(78)90447-1",
    journal = "Phys. Lett. B",
    volume = "79",
    pages = "112--113",
    year = "1978"
}

@article{Gousheh:2012pwg,
    author = "Gousheh, Siamak S. and Mohammadi, Azadeh and Shahkarami, Leila",
    title = "{An investigation of the Casimir energy for a fermion coupled to the sine-Gordon soliton with parity decomposition}",
    eprint = "1212.2089",
    archivePrefix = "arXiv",
    primaryClass = "hep-th",
    doi = "10.1140/epjc/s10052-014-3020-2",
    journal = "Eur. Phys. J. C",
    volume = "74",
    number = "8",
    pages = "3020",
    year = "2014"
}

@article{Hoffmann:2017,
author = {Hoffmann, Markus and Zimmermann, Bernd and Müller, Gideon and Schürhoff, Daniel and Kiselev, Nikolai and Melcher, Christof and Blügel, Stefan},
year = {2017},
month = {08},
pages = {308},
title = {Antiskyrmions stabilized at interfaces by anisotropic Dzyaloshinskii-Moriya interaction},
volume = {8},
journal = {Nature Communications},
doi = {10.1038/s41467-017-00313-0}
}

@article{Jackiw:1975fn,
    author = "Jackiw, R. and Rebbi, C.",
    title = "{Solitons with Fermion Number 1/2}",
    reportNumber = "MIT-CTP-517",
    doi = "10.1103/PhysRevD.13.3398",
    journal = "Phys. Rev. D",
    volume = "13",
    pages = "3398--3409",
    year = "1976"
}

@article{Jackiw:1977pu,
    author = "Jackiw, R. and Rebbi, C.",
    editor = "Shifman, Mikhail A.",
    title = "{Spinor Analysis of Yang-Mills Theory}",
    reportNumber = "MIT-CTP-619",
    doi = "10.1103/PhysRevD.16.1052",
    journal = "Phys. Rev. D",
    volume = "16",
    pages = "1052",
    year = "1977"
}

@article{Jackiw:1981ee,
    author = "Jackiw, R. and Rossi, P.",
    title = "{Zero Modes of the Vortex - Fermion System}",
    reportNumber = "MIT-CTP-928",
    doi = "10.1016/0550-3213(81)90044-4",
    journal = "Nucl. Phys. B",
    volume = "190",
    pages = "681--691",
    year = "1981"
}

@article{Kahana:1984be,
    author = "Kahana, S. and Ripka, G.",
    title = "{Baryon Density of Quarks Coupled to a Chiral Field}",
    doi = "10.1016/0375-9474(84)90692-4",
    journal = "Nucl. Phys. A",
    volume = "429",
    pages = "462--476",
    year = "1984"
}

@article{Kahana:1984dx,
    author = "Kahana, S. and Ripka, G. and Soni, V.",
    title = "{Soliton with Valence Quarks in the Chiral Invariant Sigma Model}",
    doi = "10.1016/0375-9474(84)90306-3",
    journal = "Nucl. Phys. A",
    volume = "415",
    pages = "351--364",
    year = "1984"
}

@article{Kiskis:1978tb,
    author = "Kiskis, Joe E.",
    title = "{Fermion Zero Modes and Level Crossing}",
    reportNumber = "PRINT-78-0786 (IAS,PRINCETON)",
    doi = "10.1103/PhysRevD.18.3690",
    journal = "Phys. Rev. D",
    volume = "18",
    pages = "3690",
    year = "1978"
}

@article{Kodama:2008xm,
    author = "Kodama, Yuta and Kokubu, Kento and Sawado, Nobuyuki",
    title = "{Localization of massive fermions on the baby-skyrmion branes in 6 dimensions}",
    eprint = "0812.2638",
    archivePrefix = "arXiv",
    primaryClass = "hep-th",
    doi = "10.1103/PhysRevD.79.065024",
    journal = "Phys. Rev. D",
    volume = "79",
    pages = "065024",
    year = "2009"
}

@article{Koshibae:2016aa,
author = {Koshibae, Wataru and Nagaosa, Naoto},
year = {2016},
month = {01},
pages = {10542},
title = {Theory of antiskyrmions in magnets},
volume = {7},
journal = {Nature Communications},
doi = {10.1038/ncomms10542}
}

@article{Kunz:1994ah,
    author = "Kunz, Jutta and Brihaye, Yves",
    title = "{Level crossing along sphaleron barriers}",
    eprint = "hep-ph/9403216",
    archivePrefix = "arXiv",
    doi = "10.1103/PhysRevD.50.1051",
    journal = "Phys. Rev. D",
    volume = "50",
    pages = "1051--1059",
    year = "1994"
}

@book{Manton:2004tk,
    author = "Manton, N. S. and Sutcliffe, P.",
    title = "{Topological solitons}",
    doi = "10.1017/CBO9780511617034",
    isbn = "978-0-521-04096-9, 978-0-521-83836-8, 978-0-511-20783-9",
    publisher = "Cambridge University Press",
    series = "Cambridge Monographs on Mathematical Physics",
    year = "2004"
}

@book{Manton:2022fcb,
    author = "Manton, Nicholas S.",
    title = "{Skyrmions {\textendash} A Theory of Nuclei}",
    doi = "10.1142/q0368",
    isbn = "978-1-80061-247-1, 978-1-80061-249-5",
    publisher = "World Scientific",
    month = "3",
    year = "2022"
}

@article{Nagaosa:2013ftn,
    author = "Nagaosa, Naoto and Tokura, Yoshinori",
    title = "{Topological properties and dynamics of magnetic skyrmions}",
    doi = "10.1038/nnano.2013.243",
    journal = "Nature Nanotech.",
    volume = "8",
    number = "12",
    pages = "899--911",
    year = "2013"
}

@article{Niemi:1984vz,
    author = "Niemi, A. J. and Semenoff, G. W.",
    title = "{Fermion Number Fractionization in Quantum Field Theory}",
    reportNumber = "PRINT-85-0355 (IAS,PRINCETON)",
    doi = "10.1016/0370-1573(86)90167-5",
    journal = "Phys. Rept.",
    volume = "135",
    pages = "99",
    year = "1986"
}

@article{Perapechka:2018yux,
    author = "Perapechka, I. and Sawado, Nobuyuki and Shnir, Ya",
    title = "{Soliton solutions of the fermion-Skyrmion system in (2+1) dimensions}",
    eprint = "1808.07787",
    archivePrefix = "arXiv",
    primaryClass = "hep-th",
    doi = "10.1007/JHEP10(2018)081",
    journal = "JHEP",
    volume = "10",
    pages = "081",
    year = "2018"
}

@article{Perapechka:2019upv,
    author = "Perapechka, I. and Shnir, Ya.",
    title = "{Fermion exchange interaction between magnetic Skyrmions}",
    eprint = "1901.06925",
    archivePrefix = "arXiv",
    primaryClass = "hep-th",
    doi = "10.1103/PhysRevD.99.125001",
    journal = "Phys. Rev. D",
    volume = "99",
    number = "12",
    pages = "125001",
    year = "2019"
}

@article{Perelomov:1987va,
    author = "Perelomov, A. M.",
    title = "{CHIRAL MODELS: GEOMETRICAL ASPECTS}",
    doi = "10.1016/0370-1573(87)90044-5",
    journal = "Phys. Rept.",
    volume = "146",
    pages = "135--213",
    year = "1987"
}

@article{Piette:1994,
author = {Piette, Bernard and Schroers, B. and Zakrzewski, W.},
year = {1994},
month = {06},
pages = {},
title = {Multisolitons in a Two-dimensional Skyrme Model},
volume = {65},
journal = {Zeitschrift für Physik C Particles and Fields},
doi = {10.1007/BF01571317}
}

@article{Polyakov:1975yp,
    author = "Polyakov, Alexander M. and Belavin, A. A.",
    title = "{Metastable States of Two-Dimensional Isotropic Ferromagnets}",
    journal = "JETP Lett.",
    volume = "22",
    pages = "245--248",
    year = "1975"
}

@book{Rajaraman:1982is,
    author = "Rajaraman, R.",
    title = "{SOLITONS AND INSTANTONS. AN INTRODUCTION TO SOLITONS AND INSTANTONS IN QUANTUM FIELD THEORY}",
    publisher = "North-Holland Personal Library",
    year = "1982"
}

@article{RevModPhys.51.591,
  title = {The topological theory of defects in ordered media},
  author = {Mermin, N. D.},
  journal = {Rev. Mod. Phys.},
  volume = {51},
  issue = {3},
  pages = {591--648},
  numpages = {0},
  year = {1979},
  month = {Jul},
  publisher = {American Physical Society},
  doi = {10.1103/RevModPhys.51.591},
  url = {https://link.aps.org/doi/10.1103/RevModPhys.51.591}
}

@article{Rubakov:1988aq,
    author = "Rubakov, V. A.",
    title = "{Monopole Catalysis of Proton Decay}",
    doi = "10.1088/0034-4885/51/2/002",
    journal = "Rept. Prog. Phys.",
    volume = "51",
    pages = "189--241",
    year = "1988"
}

@book{Shnir:2018yzp,
    author = "Shnir, Yakov M.",
    title = "{Topological and Non-Topological Solitons in Scalar Field Theories}",
    isbn = "978-1-108-63625-4",
    publisher = "Cambridge University Press",
    month = "7",
    year = "2018"
}

@article{Skyrme:1961vq,
    author = "Skyrme, T. H. R.",
    title = "{A Nonlinear field theory}",
    doi = "10.1098/rspa.1961.0018",
    journal = "Proc. Roy. Soc. Lond. A",
    volume = "260",
    pages = "127--138",
    year = "1961"
}

@article{Skyrme:1961vr,
    author = "Skyrme, T. H. R.",
    title = "{Particle states of a quantized meson field}",
    doi = "10.1098/rspa.1961.0115",
    journal = "Proc. Roy. Soc. Lond. A",
    volume = "262",
    pages = "237--245",
    year = "1961"
}

@article{Weinberg:1981eu,
    author = "Weinberg, Erick J.",
    title = "{Index Calculations for the Fermion-Vortex System}",
    reportNumber = "CU-TP-202",
    doi = "10.1103/PhysRevD.24.2669",
    journal = "Phys. Rev. D",
    volume = "24",
    pages = "2669",
    year = "1981"
}

@article{Zhang:2022lyz,
    author = "Zhang, Hao and Wang, Zhentao and Dahlbom, David and Barros, Kipton and Batista, Cristian D.",
    title = "{CP$^{2}$ skyrmions and skyrmion crystals in realistic quantum magnets}",
    eprint = "2203.15248",
    archivePrefix = "arXiv",
    primaryClass = "cond-mat.str-el",
    doi = "10.1038/s41467-023-39232-8",
    journal = "Nature Commun.",
    volume = "14",
    number = "1",
    pages = "3626",
    year = "2023"
}

@article{PhysRevD.92.045007,
  title = {Potentials and the vortex solutions in the $C{P}^{N}$ Skyrme-Faddeev model},
  author = "{{Y. Amari}, {P. Klimas}, {N. Sawado}, and {Y. Tamaki}, }",
  journal = {Phys. Rev. D},
  volume = {92},
  pages = {045007},
  numpages = {15},
  year = {2015},
  month = {Aug},
  publisher = {American Physical Society},
  doi = {10.1103/PhysRevD.92.045007}
}

@article{PhysRevD.100.105003,
  title = {Fermions on kinks revisited},
  author = {Klimashonok, Vladislav and Perapechka, Ilya and Shnir, Yakov},
  journal = {Phys. Rev. D},
  volume = {100},
  issue = {10},
  pages = {105003},
  numpages = {9},
  year = {2019},
  month = {Nov},
  publisher = {American Physical Society},
  doi = {10.1103/PhysRevD.100.105003},
  url = {https://link.aps.org/doi/10.1103/PhysRevD.100.105003}
}

@article{PhysRevD.101.021701,
  title = {Kinks bounded by fermions},
  author = {Perapechka, Ilya and Shnir, Yakov},
  journal = {Phys. Rev. D},
  volume = {101},
  issue = {2},
  pages = {021701},
  numpages = {5},
  year = {2020},
  month = {Jan},
  publisher = {American Physical Society},
  doi = {10.1103/PhysRevD.101.021701},
  url = {https://link.aps.org/doi/10.1103/PhysRevD.101.021701}
}

@article{PhysRevD.108.065005,
  title = {Fermion states localized on a self-gravitating non-Abelian monopole},
  author = {Dzhunushaliev, Vladimir and Folomeev, Vladimir and Shnir, Yakov},
  journal = {Phys. Rev. D},
  volume = {108},
  issue = {6},
  pages = {065005},
  numpages = {10},
  year = {2023},
  month = {Sep},
  publisher = {American Physical Society},
  doi = {10.1103/PhysRevD.108.065005},
  url = {https://link.aps.org/doi/10.1103/PhysRevD.108.065005}
}

@article{PhysRevD.111.084072,
  title = {Euclidean AdS wormholes and gravitational instantons in the Einstein-Skyrme theory},
  author = {Canfora, Fabrizio and Corral, Crist\'obal and Diez, Borja},
  journal = {Phys. Rev. D},
  volume = {111},
  issue = {8},
  pages = {084072},
  numpages = {13},
  year = {2025},
  month = {Apr},
  publisher = {American Physical Society},
  doi = {10.1103/PhysRevD.111.084072},
  url = {https://link.aps.org/doi/10.1103/PhysRevD.111.084072}
}

@article{PhysRevLett.100.047203,
  title = {Pairing of Solitons in Two-Dimensional $S=1$ Magnets},
  author = {Ivanov, B. A. and Khymyn, R. S. and Kolezhuk, A. K.},
  journal = {Phys. Rev. Lett.},
  volume = {100},
  issue = {4},
  pages = {047203},
  numpages = {4},
  year = {2008},
  month = {Jan},
  publisher = {American Physical Society},
  doi = {10.1103/PhysRevLett.100.047203},
  url = {https://link.aps.org/doi/10.1103/PhysRevLett.100.047203}
}

\end{document}